\documentclass[preprint,12pt]{elsarticle}

\usepackage{amssymb}
\usepackage{color,xcolor}
\usepackage{amsmath}
\usepackage{mathrsfs}

\newcommand\beq{\begin{equation}}
\newcommand\eeq{\end{equation}}

\newcommand{\ben}{\begin{eqnarray}}
	\newcommand{\een}{\end{eqnarray}}
\newcommand{\benn}{\begin{eqnarray*}}
	\newcommand{\eenn}{\end{eqnarray*}}

\newcommand{\apar}{ A_{\parallel}}

\newcommand{\bpar}{B_z}
\newcommand{\pa}{\partial}

\newcommand{\lapp}{\Delta_\perp}

\newcommand{\gpar}{\nabla_{\parallel}}

\newcommand{\bcdot}{\mathbf{\cdot}}

\newcommand{\nno}{\nonumber}

\newcommand{\bdep}{\frac{\beta_e}{2} d_e^2}
\newcommand{\gamue}{G_{10e}}
\newcommand{\gamde}{G_{20e}}
\newcommand{\gamui}{G_{10i}}
\newcommand{\gamdi}{G_{20i}}

\journal{Fundamental Plasma Physics}

\begin{document}

\begin{frontmatter}



\title{Gyrofluid simulations of turbulence and reconnection\\ 
in space plasmas}


\author{T. Passot$^{a}$, S. S. Cerri$^{a}$, C. Granier$^{b}$, D. Laveder$^{a}$, P.L. Sulem$^{a}$, E. Tassi$^{a}$}
\affiliation{organization={Université Côte d'Azur, Observatoire de la Côte d'Azur,  CNRS, Laboratoire J.L. Lagrange},
            addressline={Boulevard de l'Observatoire, CS  34229}, 
            city={06304 Nice Cedex 4},
            country={France},}
\affiliation{organization={Max Planck Institute for Plasma Physics}, 
            adressline={Boltzmannstraße 2}, 
            city={85748 Garching}, 
            country = {Germany}}

\begin{abstract}
A Hamiltonian two-field gyrofluid model is used to investigate the dynamics of  an electron-ion collisionless plasma subject to a strong ambient magnetic field, within a spectral range extending from the magnetohydrodynamic (MHD) scales to the electron skin depth. This model isolates Alfv\'en,  Kinetic Alfv\'en and Inertial Kinetic Alfv\'en waves that play a central role in space plasmas, and extends standard  reduced fluid models to broader ranges of the plasma parameters.  Recent numerical results are reviewed, including (i) the reconnection-mediated MHD turbulence developing  from the collision of counter-propagating Alfv\'en wave packets, (ii) the specific features of the cascade dynamics in strongly imbalanced turbulence, including a possible link between the existence of a spectral transition range and the presence of co-propagating wave interactions at sub-ion scales, for which new simulations are reported, (iii) the influence of the ion-to-electron temperature ratio in two-dimensional collisionless magnetic reconnection. The role of electron finite Larmor radius corrections is  pointed out and the extension of the present model to a four-field gyrofluid model is discussed. Such an extended model accurately describes electron finite Larmor radius effects at small or moderate values of the electron beta parameter, and also retains the coupling to slow magnetosonic waves.
\end{abstract}


\begin{highlights}
\item Research highlight 1
\item Research highlight 2
\end{highlights}


\end{frontmatter}








\section{Introduction}
\label{sec:intro}
The description of turbulence and reconnection in collisionless space plasmas such as the solar wind or planetary magnetospheres is challenging, especially when a wide range of scales is to be retained, or when parametric studies are to be performed. In such regimes, the huge computational resources required by fully kinetic simulations are indeed at the limit, and often beyond, the capabilities of today's computers. 
Both in astrophysical environments and in fusion devices,  when the plasma is subject to a strong ambient field and the focus is on the 
low-frequency dynamics, an efficient tool is provided by gyrofluid models which, unlike usual fluid approaches, take into account some important kinetic effects. 
Due to their ability to capture the relevant physics at a reasonable computational cost, gyrofluid models and their fluid-scale reduction (namely, reduced-MHD) have been increasingly employed during the past decade  in numerical simulations aimed at investigating the properties of Alfv\'enic turbulence in space plasmas, in various ranges of scales,  from the fluid ones to the kinetic ones \citep[e.g.,][]{PerezBoldyrevAPJL2010,ChenMNRAS2011,Perez12,Perez13,Boldyrev-Perez12,Chen-Bold17,Meyrand21,Borgogno22,ZhouMNRAS2023}.

The aim of this  paper is to present a short review of recent results, together with some new developments  obtained by simulations  based on a two-field Hamiltonian gyrofluid model,  introduced in \cite{PST18}. This model retains ion finite Larmor radius (FLR) corrections, parallel magnetic field fluctuations and electron inertia, in the regime of small $\beta_e$ (where $\beta_e$ indicates the ratio between equilibrium electron kinetic pressure and guide-field magnetic pressure), and arbitrary ion-to-electron temperature ratio.  The model provides, within a unique system of equations, a description of  the quasi-perpendicular dynamics of Alfv\'en waves (AWs), kinetic-Alfv\'en waves (KAWs), as well as of inertial kinetic-Alfv\'en waves (IKAWs), such as those detected by MMS in the Earth’s magnetosheath \cite{Chen-Bold17}. 
Moreover, this model is  suitable for  studying collisionless reconnection at the electron scale, as observed in the Earth's turbulent magnetosheath \cite{Phan18}. 

The paper is organized as follows. After describing the two-field model in Section \ref{sec:model}, we highlight in Section \ref{sec:RMHD_regime} the development of a reconnection-mediated regime in the three-dimensional reduced-MHD (RMHD) turbulence  that develops from collisions of counter-propagating Alfv\'en-wave packets, at small to moderate levels of the nonlinearity parameter. In  Section \ref{sec:cascades}, we discuss Alfvenic turbulent cascades that develop from the MHD to the sub-ion scales, when there is a strong imbalance between the energies of the co- and counter-propagating waves (such as observed by the Parker Solar Probe (PSP) close to the Sun \cite{Chen2020}). In this regime, a so-called “helicity barrier” \cite{Meyrand21} emerges between the MHD and the kinetic scales, which provides a mechanism for the development of the transition range in the magnetic-field spectrum across the ion scales.  Such a transition was first reported from Cluster observations in the solar wind (see e.g \cite{Sahraoui10,Alexandrova13}) and more recently from PSP data \cite{Bowen-PRL20,Bowen23}. The origin of the transition zone at 1 AU observed by Cluster is not completely clear. It is possibly associated with ion Landau damping \cite{Sahraoui10} and appears to be sensitive to the power of fluctuations within the inertial range \cite{Bruno14}. Differently, the PSP observations made in the inner heliosphere indicate that the slope becomes steeper for increasing normalized cross helicity (a quantity that measures the imbalance between outward and inward propagating fluctuations) \cite{Huang_2021}, in line with the gyrofluid simulations reported in \cite{PSL22}.
We also show that, in regimes of weak nonlinearities, where the parallel dissipation remains negligible, the effect of co-propagating wave interactions is to induce a steeper magnetic spectrum in this range. In Section \ref{sec:reconnection}, we discuss, as a function of the ratio of ion to electron temperatures, the development of turbulence that results from secondary instabilities in the nonlinear phase of two-dimensional (2D) collisionless reconnection. It turns out that the  nature of this turbulence, which transfers energy to sub-electron scales, and  leads to the generation of small-scale vortices, is affected by the presence of electron  FLR corrections. This effect is studied in more detail in Section \ref{sec:limitation} in the framework of homogeneous two-dimensional turbulence. We conclude by presenting a new Hamiltonian four-field model that provides an accurate description of the sub-electronic scales and is not limited to small values of $\beta_e$. It also includes the coupling to slow-magnetosonic waves and can thus describe the decay instability at the MHD scales. Section \ref{sec:conclusion} summarizes these findings and provides a brief discussion of forthcoming developments amenable to gyrofluid descriptions. A new derivation of the two-field model from the four-field gyrofluid model, is presented in the Appendix.

\section{A two-field Hamiltonian gyrofluid model}\label{sec:model}

\subsection{Formulation and main properties of the model}
Gyrofluid models, which are derived from a gyrokinetic model, involve the gyrokinetic scaling (see e.g. \citet{HCD06}) and thus  prescribe an anisotropic dynamics with transverse scales much smaller than the parallel ones and frequencies small compared to the ion gyrofrequency. Fast magnetosonic waves are thus averaged out.  

The two-field reduced gyrofluid model considered in this paper was derived in \cite{PST18} as a restriction of the  model of \citet{Bri92}. An alternative derivation is presented in the Appendix. The model isolates the  Alfv\'en wave dynamics and describes small perturbations of a homogeneous equilibrium state characterized by a  density $n_0$, isotropic ion and electron temperatures $T_{0i}$ and $T_{0e}$, and subject to a strong ambient magnetic field of amplitude $B_0$ along the $z$-direction.
The main plasma characteristic spatial scales (inertial lengths $d_r=v_A/\Omega_r$ and Larmor radii $\rho_r =v_{th\,r}/\Omega_r$) for the ions ($r=i$) and electrons ($r=e$), are defined in the form
\begin{equation}
	d_i = \sqrt{\frac{2}{\beta_e}} \rho_s , \quad
	d_e= \sqrt{\frac{2}{\beta_e}}\delta \rho_s , \quad
	\rho_i = \sqrt{2\tau} \rho_s , \qquad  \rho_e= \sqrt{2} \delta \rho_s,
\end{equation}
where $\beta_e= 8 \pi n_0 T_{0e}/B_0^2$, $\tau = T_{0i}/T_{0e}$, $\delta^2= m_e/m_i$ is
the electron to ion mass ratio, $v_{th \, r}=(2T_r/m_r)^{1/2}$ are the particle thermal velocities,  and $v_A=B_0/(4\pi n_0 m_i)^{1/2}
=c_s \sqrt{2/\beta_e}$ is the Alfv\'en speed.
We normalize lengths by the  
sonic Larmor radius $\rho_s  = c_s/\Omega_i$, where $c_s = \sqrt {T_{0e}/m_i}$ is the so-called ion-sound speed, 
time by the inverse ion gyrofrequency $\Omega_i= eB_0/(m_i c)$, the parallel magnetic fluctuations $B_z$ by  $B_0$, the electron gyrocenter density $N_e$ by $n_0$, the electric potential $\varphi$ by $T_e/e$ and the parallel magnetic potential $A_\|$ by  $B_0\rho_s$. The equations for $N_e$ and $A_\|$ then read 
\begin{eqnarray}
	&&\partial_t N_e +[\varphi,N_e]-[B_z,N_e]+\frac{2}{\beta_e}\nabla_\| \Delta_\perp A_\|=0\label{eq:gyro-2fields-Ne} \label{eq:Ne}\\
	&&\partial_t \left(1-\frac{2\delta^2}{\beta_e}\Delta_\perp \right)A_\| -\left[\varphi,\frac{2\delta^2}{\beta_e}\Delta_\perp A_\| \right]
	+\left[B_z,\frac{2\delta^2}{\beta_e}\Delta_\perp A_\| \right] 
	+ \nabla_\| (\varphi-N_e-B_z)=0,
 \label{eq:A}\nonumber\\
\end{eqnarray}
with $B_z$ and  $\varphi$ given by 
\begin{eqnarray}
	&&\left (\frac{2}{\beta_e}  +(1+2\tau)(\Gamma_0 - \Gamma_1) \right ) B_z=
	\left ( 1 -\frac{\Gamma_0-1}{\tau} -\Gamma_0
	+\Gamma_1  \right)\varphi  \label{gyro:Bzphi}\\
	&&N_e=\left ( \frac{\Gamma_0-1}{\tau}  +\delta^2\Delta_\perp\right )\varphi
	-(1-\Gamma_0+\Gamma_1) B_z.\label{eq:gyro-Ne-phi}
\end{eqnarray}
Equations (\ref{eq:Ne}) and (\ref{eq:A}) correspond to the continuity equation for the electron gyrocenter density fluctuations $N_e$ and to Ohm's law, respectively. The relations (\ref{gyro:Bzphi}) and (\ref{eq:gyro-Ne-phi}), on the other hand, express the quasi-neutrality condition and the perpendicular component of Amp\`ere's law, respectively. 

The operator $\Delta_\perp = \partial_{xx} + \partial_{yy}$ denotes the Laplacian in the plane transverse to the ambient field and  $[f,g]= \partial_x f \partial_y g-\partial_y f \partial_x g$ is the canonical bracket of two scalar functions $f$ and $g$.
The (non-local) operator $\Gamma_n$ is associated with  the Fourier multiplier  $\Gamma_n(\tau k_\perp^2)$, defined by  $\Gamma_n(x) = I_n(x) e^{-x}$ where $I_n$ is the modified Bessel function of first type of order n. For a scalar function $f$, the parallel gradient operator $\gpar$ is defined by $\gpar f=-[\apar , f]+\frac{\pa f}{\pa z}$.

Writing $ B_z = M_1 \varphi$, 
and $N_e = -M_2 \varphi$,   
$B_z$ and $\varphi$  can be expressed in terms of $N_e$.  In the previous sentence we indicated with $M_1$ and $M_2$ two operators, (with $M_2$, in particular,  invertible) the expression of which can be found explicitly in Fourier space from Eqs. (\ref{gyro:Bzphi}) and (\ref{eq:gyro-Ne-phi}).

Quasineutrality relates the ion and electron particle number densities $n_i$ and $n_e$ by
$n_i=n_e=N_e+B_z$, the latter equality being only valid in the absence of electron inertia and FLR contributions.

In addition to ion FLR corrections and parallel magnetic fluctuations, \textcolor{black}{the model} retains electron inertia as well as an electron FLR contribution which becomes relevant when the ion-electron temperature ratio $\tau$ is comparable to or larger than the inverse electron beta  $1/\beta_e$.  Furthermore, this
model possesses  a noncanonical Hamiltonian structure. It covers a spectral
range extending from the MHD scales (large compared to 
$d_i$) to scales comparable to $d_e$ but nevertheless large compared to the electron Larmor radius $\rho_e$ (in order to prevent the full FLR electron corrections to be relevant). The latter condition requires that  $\rho_e/d_e=\beta_e^{1/2}$ be small enough, a regime where Landau damping can efficiently homogenize electron temperatures along the magnetic field lines. The present model thus assumes isothermal electrons, which is a good approximation when neglecting dissipation phenomena \citep{TSP16,Sulem2016}.

Equations (\ref{eq:Ne})--(\ref{eq:A}) preserve two quadratic invariants, namely the energy 
\begin{equation}
{\mathcal E} = \frac{1}{2} \int \Big ( \frac{2}{\beta_e} |\nabla_\perp A_\| |^2 
+ \frac{4\delta^2}{\beta_e^2}|\Delta_\perp A_\| |^2 
- N_e(\varphi -N_e-B_z) \Big ) d^3 {x}, \label{energy}
\end{equation}
and the generalized cross helicity (GCH)
\begin{equation}
{\mathcal C} =-\int N_e \Big (1 - \frac{2\delta^2}{\beta_e} \Delta_\perp\Big) A_\|d^3 {x}.\label{defC}
\end{equation}
At large scales, ${\mathcal C}$ reduces, up to a sign, to the MHD cross helicity, while at small scales, it identifies (up to electron-inertia corrections) with the parallel contribution to the magnetic helicity.

\subsection{A few limiting regimes}
Equations (\ref{eq:Ne})-(\ref{eq:A}) reduce to classical systems in various limits (see \cite{PS19} for more details). 
\begin{itemize}
 \item{\bf RMHD \& Hall-RMHD ($\tau k^2_\perp\ll 1$, $\tau\ll 1$, $\beta_e\lesssim 1$, $\delta=0$)}\\
 In this limit, the gyro-fluid equations reduce to the equations for dispersive Alfvén waves, i.e. the reduced-MHD regime with the Hall effect. This is a two-field reduction of the HRMHD model of \citet{Schekochihin09} when neglecting the parallel ion velocity $u_i$ and assuming that $B_z$ is prescribed by the instantaneous potential $\varphi$. When considering even larger scales (i.e., much larger that the ion-inertial length $d_i$), the system further reduces to the 2-field RMHD.

\item{\bf Schep et al. model ($\tau k^2_\perp\ll 1$,  $\tau\ll 1$, $\beta_e\ll 1$, $\delta\neq0$)}\\
With the same conditions on $\tau$ and $\tau k^2$ as for the HRMHD regime, but restricting to $\beta_e\ll 1$ and allowing for $\delta \ne 0$, one easily recovers a two-field, cold-ion version of the model of \citet{Schep94}, which was extensively used for studying collisionless reconnection (see, for instance \cite{Caf98,Bet22}).

\item{\bf Isothermal KREHM  limit ($\tau k^2_\perp \sim 1$, $\tau \sim  1$, $\beta_e \sim \delta^2 $,  $\delta \ne 0$)}\\
In this case, our gyrofluid equations recover the isothermal electron limit of the kinetic-reduced electron heating model (KREHM) derived in \citet{Zocco11}. 

\item{\bf Electron-RMHD ($\tau k^2_\perp\gg 1$, $\tau \sim  1$, $\beta_e\lesssim 1$,  $\delta =0$)}\\
In this case, the model reproduces the so-called electron-reduced-MHD (ERMHD) regime for KAWs \citep{Schekochihin09}.

\item{\bf Inertial-KAW regime ($\tau k^2_\perp\gg 1$, $\tau\gg 1$, $\beta_e\lesssim 1$, $\delta\ne 0$)}\\
Interestingly, when $\tau\gg 1$, together with $\tau k^2_\perp\gg 1$, $\beta_e\lesssim 1$ and $\delta\ne 0$, the limit of inertial kinetic Alfvén waves (IKAWs) \citep{Chen-Bold17, PST17} is recovered. The system reads
	\begin{align}
		&\partial_t \left(1-\frac{2\delta^2}{\beta_e}\Delta_\perp \right) A_\| -\left[\varphi, \frac{2\delta^2}{\beta_e}\Delta_\perp A_\| \right]+ \nabla_\|  \varphi=0 \label{eq:IKAW1}\\
		&\partial_t \left (1+\frac{2}{\beta_i} -\frac{2\delta^2}{\beta_e}\Delta_\perp \right)\varphi - \left[\varphi,\frac{2\delta^2}{\beta_e}\Delta_\perp \varphi \right]- \frac{4}{\beta_e^2} \nabla_\|\Delta_\perp A_\| =0, \label{eq:IKAW2}
	\end{align}
where $\beta_i=\tau\beta_e$ refers to the ion beta parameter. In Eq. (\ref{eq:IKAW2}), the term 
$\partial_t \left ( \frac{2\delta^2}{\beta_e}\Delta_\perp \varphi \right )$, which in this regime arises at the dominant order, originates from the  $\delta^2 \Delta_\perp \varphi$ contribution in Eq. (\ref{eq:gyro-Ne-phi}), which is a lower-order electron FLR correction and can be neglected when $\tau \simeq 1$, but has to be kept when $\tau \gg 1$ because, in this limit,  its magnitude becomes comparable to that of the other terms. Note however that in the latter case,  the retained electron FLR correction also leads to subdominant terms in Eq. (\ref{eq:A}). 

\end{itemize}

\subsection{Alternative normalization of the model equations}

The model (\ref{eq:Ne})-(\ref{eq:gyro-Ne-phi}) is formulated in \cite{GTLPS23} with a normalization based on the Alfv\'en time, thus different with respect to the original reference \cite{PST18}. This helped in the comparison with several studies present in the literature, concerning in particular a two-field reduction of the model of \citet{Schep94}. Therefore, we find it appropriate to also present the gyrofluid model with this alternative normalization, which will also be adopted in Sec. \ref{sec:reconnection} and in the Appendix. In this framework, the model (\ref{eq:Ne})-(\ref{eq:gyro-Ne-phi}) reads
\begin{align}
&\frac{\pa N_e}{\pa t}+ [\varphi - \rho_s^2 B_z , N_e ] +\nabla_\| \lapp \apar=0, \label{contfin}\\
&\frac{\pa }{\pa t}(1 - d_e^2 \lapp )\apar-[\varphi , d_e^2 \lapp \apar]+\rho_s^2[B_z , d_e^2 \lapp \apar]\nonumber\\
& + \nabla_\| (\varphi - \rho_s^2 N_e - \rho_s^2 B_z)=0, \label{ohmfin}\\
& \left( \frac{2}{\beta_e} + (1 + 2 \tau) (\Gamma_{0}  -\Gamma_{1}) \right) \bpar=\left(1 - \frac{\Gamma_{0}  -1}{\tau} -\Gamma_{0}  +\Gamma_{1}  \right) \frac{\varphi}{\rho_s^2}, \label{ampperpfin}\\
&N_e =   \left(\frac{\Gamma_{0} -1}{\tau}+\bdep \lapp\right)\frac{\varphi}{\rho_s^2}-(1 - \Gamma_{0}  +\Gamma_{1}) B_z, \label{qnfin}
\end{align}
and the normalization is given by
\begin{align}
&t=\frac{v_A}{L}\hat{t}, \qquad x=\frac{\hat{x}}{L}, \qquad y=\frac{\hat{y}}{L}, \qquad z=\frac{\hat{z}}{L},  \nonumber \\
& d_e=\frac{\hat{d}_e}{L}, \qquad \rho_s=\frac{\hat{\rho}_s}{L}, \label{norma}\\
&N_e=\frac{L}{\hat{d}_i}\frac{\hat{N}_e}{n_0}, \qquad \varphi=\frac{c}{v_A}\frac{\hat{\varphi}}{L B_0}, \qquad \apar=\frac{\hat{A}_\parallel}{L B_0}, \qquad \bpar=\frac{L}{\hat{d}_i}\frac{\hat{B}_z}{B_0},\nonumber
\end{align}
where the hats denote dimensional quantities and $L$ is a characteristic length of variation in the perpendicular plane. With this normalization, the operator $\Gamma_n$ is associated with the Fourier symbol
$I_n (\tau \rho_s^2 k_\perp^2) \exp(-\tau \rho_s^2 k_\perp^2)$.

The gyrokinetic/gyrofluid ordering involves a small parameter $\varepsilon=L/L_\|$, where $L_\|$ is the characteristic length scale along the guide field. This parameter, which also measures the typical field amplitudes and frequencies, can be scaled out in the final gyrofluid equations, making all quantities of order unity. This choice is made in the following sections.

\section{Reconnection-mediated turbulence in the RMHD regime without imbalance: the weakly nonlinear scenario }
\label{sec:RMHD_regime}

In this Section, we first give an overview of the ideas underlying the theory of the turbulent cascade at large (MHD) scales, from the concept of a critically balanced cascade to a regime mediated by reconnection processes (``tearing instability''). Then, we highlight some recent numerical evidences of such a regime, with a focus on a work employing the gyrofluid model (\ref{eq:gyro-2fields-Ne})-(\ref{eq:gyro-Ne-phi}).

\subsection{Alfv\'enic turbulence at fluid scales without imbalance: from critical balance to the tearing-mediated regime, through dynamic alignment}\label{subsec:MHDturb_overview}

In the presence of a background mean magnetic field $\boldsymbol{B}_0$, the cascade of Alfv\'enic fluctuations is naturally anisotropic with respect to its direction already at ``large'' (i.e., MHD) scales. 
In this range of scales, the typical wave-vector component of the fluctuations  parallel to the local mean field (which involves scales significantly larger than those associated with the considered wave vector) tend to be much smaller than the perpendicular one, i.e., $k_\| \ll k_\perp$ (for the purpose of this subsection, we use the physical variables).
If critical balance (CB) between the linear and nonlinear timescales of the fluctuations can be assumed, \citet{GS95} originally predicted a perpendicular power spectrum $\propto\,k_\perp^{-5/3}$ and a wave-vector anisotropy $k_\|\propto k_\perp^{2/3}$, corresponding to a spectrum $\propto\,k_\|^{-2}$ in the field-parallel direction. 
Still within such CB scenario, \citet{BoldyrevPRL2006} suggested that turbulent fluctuations would undergo scale-dependent alignment/anti-alignment in the field-perpendicular plane due to the continuous shearing produced by the interaction of counter-propagating AW packets; this process has been known as ``dynamic alignment'', and is such that the angle $\theta_{k_\perp}$ between magnetic and kinetic turbulent fluctuations at wavenumber $k_\perp$ exhibits a scaling $\sin\theta_{k_\perp}\propto k_\perp^{-1/4}$. The consequence of this effect is to produce 3D-anisotropic turbulent eddies and a cascade whose spectrum scales as $k_\perp^{-3/2}$, where the spectral anisotropy now follows a $k_\|\propto k_\perp^{1/2}$ relation (here $k_\perp$ is the wave vector associated to the shortest length-scale $\lambda$ of these three-dimensional eddies, and is perpendicular to both the direction of the mean magnetic field and of magnetic-field fluctuations); the parallel spectrum in this case still scales as $k_\|^{-2}$.

The 3D nature of the eddies when dynamic alignment occurs has fundamental implications for the development of current sheets and magnetic reconnection. In fact, it has been known since long time that the formation of current sheets (CSs) is another fundamental aspect of the turbulent cascade in plasmas~\citep[e.g.,][]{PolitanoPOP1995,BiskampMuellerPOP2000,ZhdankinAPJ2013,MakwanaPOP2015,SistiAA2021}. In most cases, the CSs that are formed in this way are tearing-unstable, i.e., they are disrupted by magnetic reconnection~\citep[e.g.,][]{ServidioNPG2011,ZhdankinAPJ2015,AgudeloRuedaJPP2021}. These reconnection processes are so ubiquitous in turbulence that have been suggested to possibly mediate the energy transfer at both MHD~\citep[e.g.,][]{CarbonePOF1990,BoldyrevLoureiroAPJ2017, MalletMNRAS2017,ComissoAPJ2018,TeneraniVelli2020} and kinetic~\citep[e.g.,][]{CerriCalifanoNJP2017,FranciAPJL2017,LoureiroBoldyrevAPJ2017,MalletJPP2017} scales. When this transfer at MHD scales is mediated by reconnection, the resulting turbulence is said to be in the ``tearing-mediated'' regime, and it is characterized by a steep $k_\perp^{-11/5}$ spectrum. This regime naturally emerges in a critically balanced cascade only when dynamic alignment occurs, and it develops starting from a transition scale $\lambda_*\sim (k_\perp^*)^{-1}$ when the following conditions are satisfied: (i) the 3D-anisotropic turbulent eddies set up a tearing-unstable configuration in the plane perpendicular to the mean magnetic field, and (ii) this configuration evolves slowly enough to allow the tearing instability to grow and disrupt it (namely, when $\gamma_{\lambda}^{\rm t}\tau_{\rm nl,\lambda}\gtrsim 1$, where $\gamma_{\lambda}^{\rm t}$ is the tearing-instability growth rate at perpendicular scale $\lambda$ and $\tau_{\rm nl,\lambda}$ is the eddy turnover time at the same scale). 
The latter condition depends upon the plasma properties (e.g., resistivity), while the former is a direct consequence of the scale-dependent alignment of turbulent fluctuations (a process that needs to occur only up to a scale where tearing instability grows sufficiently fast to mediate the transfer; below such scale, reconnection will anyway tend to disrupt such alignment, at least locally).

Finally, let us take a step back and consider the case in which critical balance is not initially satisfied at the injection scales. When the injection regime is such that the large-scale nonlinear parameter, which measures the ratio between the Alfv\'en time $\tau_{\rm A,0}\sim(k_{\|,0}\,v_{\rm A,0})^{-1}$ and the nonlinear timescale $\tau_{\rm nl,0}\sim(k_{\perp,0}\,\delta b_{\perp,0})^{-1}$ at injection scales (measuring magnetic-field fluctuations in Alfv\'enic units $\delta b_{\perp}=\delta B_\perp/\sqrt{4\pi\varrho_0}$, with $\varrho_0$ indicating the equilibrium mass density), is significantly smaller  than unity (i.e., $\chi_0\sim (k_{\perp,0}/k_{\|,0})(\delta b_{\perp,0}/v_{\rm A,0})=(k_{\perp,0}/k_{\|,0})(\delta B_{\perp,0}/B_0)<1$), then the Alfv\'enic cascade develops in the so-called ``weak regime''. In this regime, the turbulent transfer does not develop a parallel cascade (i.e., $k_\|\sim k_{\|,0}\sim {\rm const.}$), and only proceeds through perpendicular scales, developing a perpendicular cascade with a $k_\perp^{-2}$ spectrum \citep{NgBhattacharjeePOP1997,Galtier00}. However, due to its own fluctuations' scaling, an initially weak cascade cannot proceed to arbitrary small scales, as the scale-dependent nonlinear parameter increases with decreasing scale as $\chi_{k_\perp}\propto k_\perp^{1/2}$. As a result, there is a scale $\lambda_{\perp,{\rm CB}}\sim k_{\perp,{\rm CB}}^{-1}\propto(\delta B_{\perp,0}/B_0)^2$ at which the nonlinear parameter becomes of order unity, $\chi_{k_{\perp,{\rm CB}}}\sim 1$, and the cascade transitions to the critically balanced regime described above. In this weak-cascade scenario, however, dynamic alignment of fluctuations has not been included in the picture. In \citet{CerriAPJ2022} this assumption has been relaxed, and two different regimes with new scaling have been proposed in a scenario where weakly-nonlinear turbulence can undergo a scale-dependent alignment, namely a moderately weak, dynamically aligned cascade with a $k_\perp^{-3/2}$ spectrum and an asymptotically weak regime ($\chi_0\ll1$) in which dynamic alignment produces a $k_\perp^{-1}$ spectrum. Some interesting consequences of this new scenario is that: (i) dynamic alignment in the weak regime allows for a transition to a tearing-mediated cascade at scales much larger that the one predicted by a critically balanced case, and (ii) in the $\chi_0\ll1$ regime, the cascade would never reach critical balance due to its own scaling, and only a tearing-mediated regime can eventually emerge and allow for a transition to strong turbulence. These scalings and the alignment of fluctuations at weak nonlinearities were indeed shown to occur by means of high-resolution numerical simulations employing the gyro-fluid model of Section~\ref{sec:model} in the RMHD regime (see next Section~\ref{subsec:rec-mediated-turb_simulations}).

\begin{figure}
    \centering
    \includegraphics[width=0.95\textwidth]{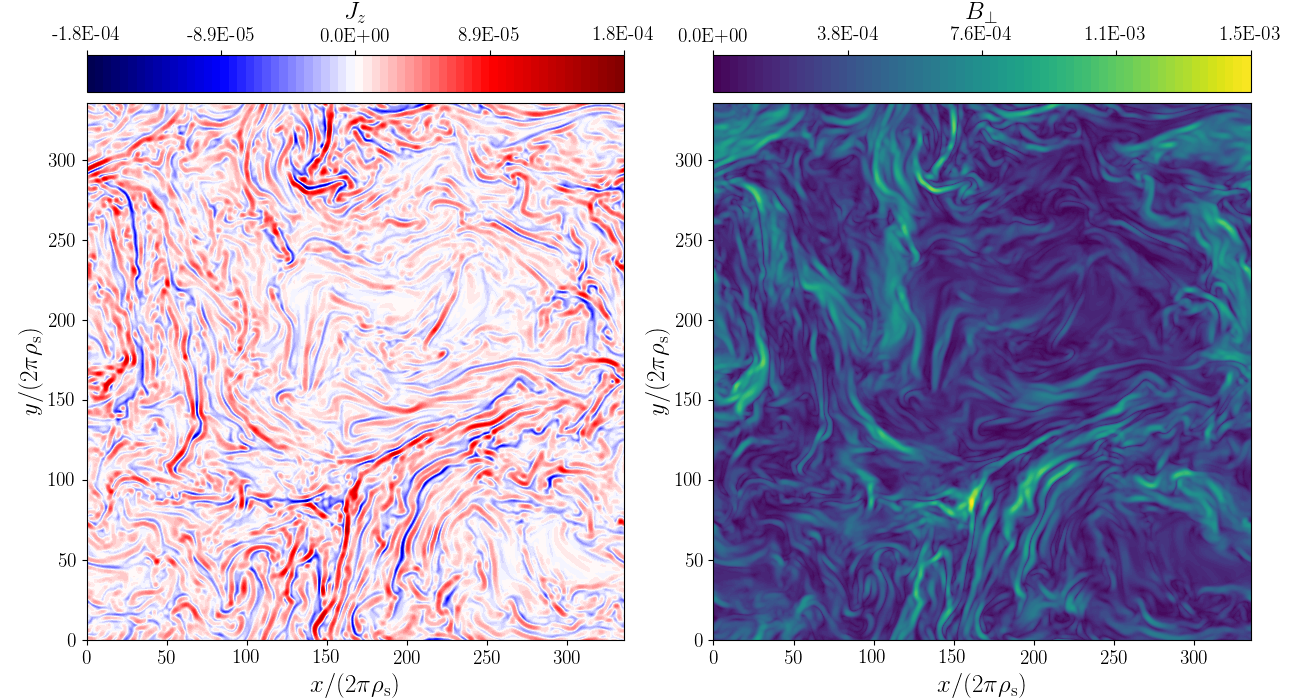}\\
	\includegraphics[width=0.95\textwidth]{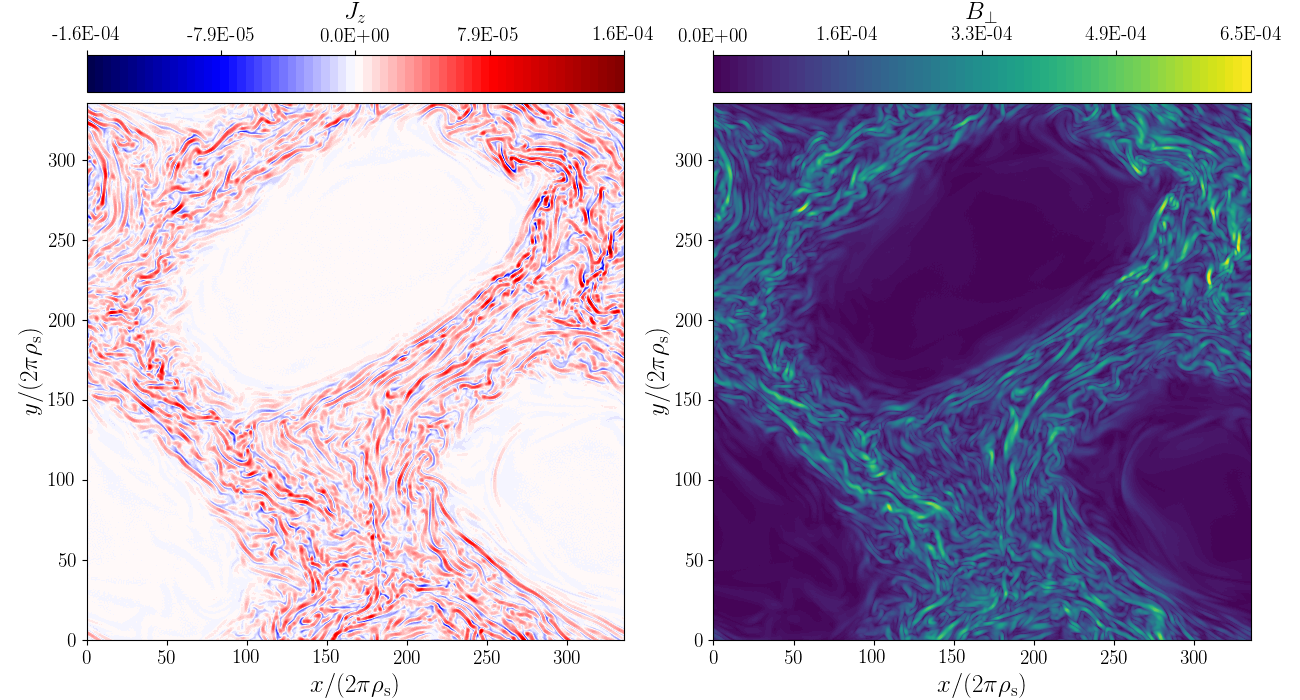}\\
	\includegraphics[width=0.95\textwidth]{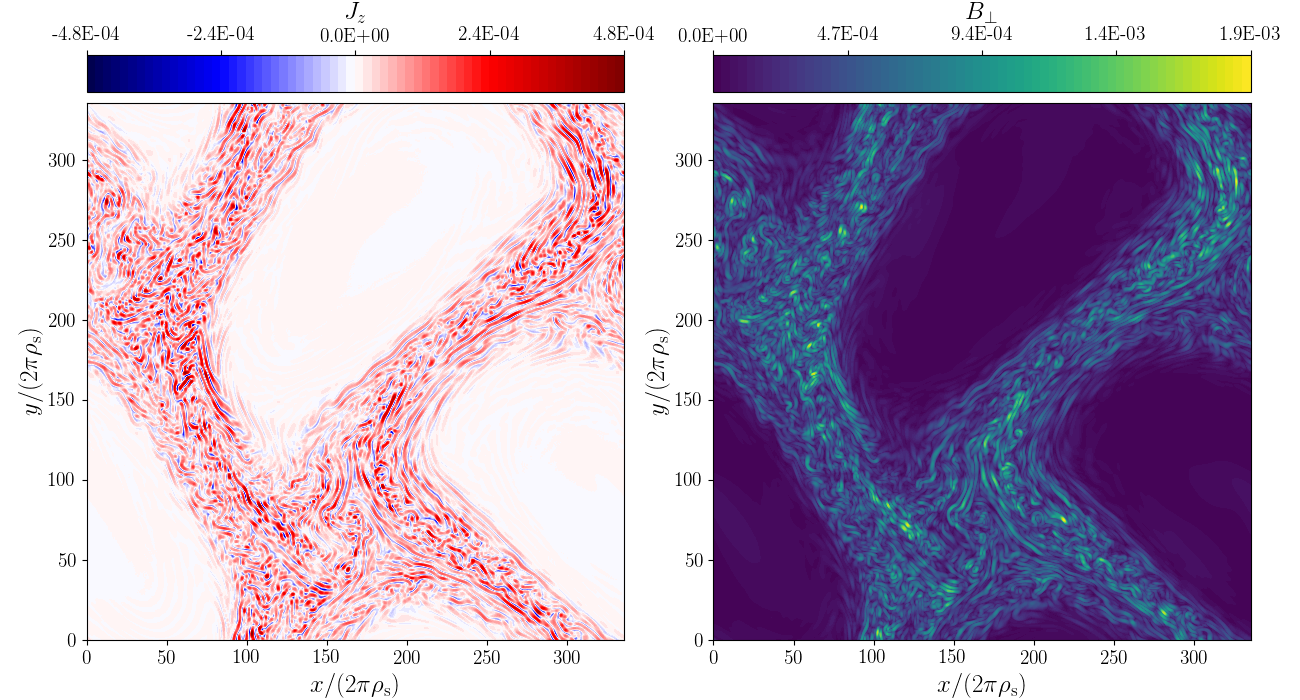}
	\caption{Parallel current density $J_z$ (left) and perpendicular magnetic field $B_\perp$ (right) in the $xy$-plane at $z=L_z/2$, during the turbulent state, for different nonlinearity regimes: $\chi_0\simeq1$ (top row), $\chi_0\simeq0.5$ (middle row), and $\chi_0\simeq0.1$ (bottom row). In all the figures, the length unit is specified for the sake of clarity.}
	\label{fig:AWcoll_contours}
\end{figure}

\subsection{Tearing-mediated range in Alfv\'enic turbulence from simulations}\label{subsec:rec-mediated-turb_simulations}

A steeper tearing-mediated range at fluid scales (i.e., well above any plasma micro-scale) was clearly shown to emerge from a larger-scale turbulent cascade via 2D simulations~\citep{DongPRL2018}.
Only very recently, the existence of such tearing-mediated regime was proven to exist also in fully 3D simulations by \citet{CerriAPJ2022} and, simultaneously, by \citet{DongSCIADV2022}.
However, the two works adopted different approaches in different regimes. 
In \citet{DongSCIADV2022}, a MHD model with broad-band injection of critically balanced fluctuations was employed; this approach required a really huge numerical resolution in order to meaningfully separate the reconnection scale $\lambda_*$ from the actual dissipation scales and obtain a $-11/5$ spectrum over a modest range of $k_\perp$ (roughly a factor $\sim3$; see their Fig.2). 
\begin{figure}
    \centering
    \includegraphics[width=0.9\textwidth]{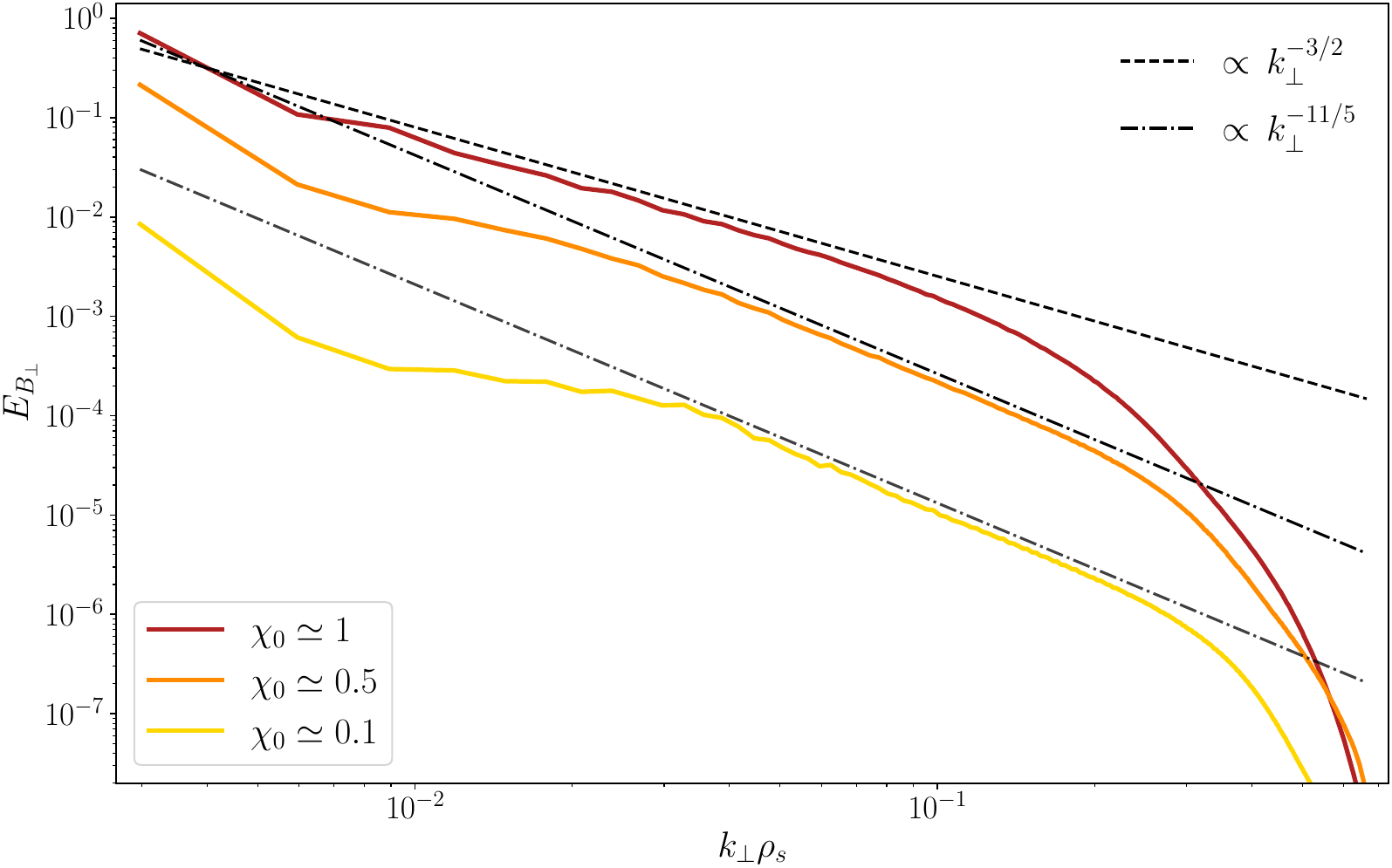}
	\caption{Energy spectra of $B_\perp$ fluctuations versus $k_\perp\rho_{\rm s}$ arising from collisions of Alfv\'en-wave packets at different large-scale nonlinearities $\chi_0$ (see legend). Spectra are averaged over the quasi-steady turbulent state, and relevant power laws are shown as a reference.}
	\label{fig:AWcoll_spectra}
\end{figure}
In \citet{CerriAPJ2022}, the problem was addressed from the point of view of the building blocks of the Alfv\'enic cascade, i.e., counter-propagating Alfv\'en-wave (AW) packets, and within the Hamiltonian two-field gyro-fluid  model presented in Section~\ref{sec:model}. The simulations were performed in a regime $\beta_{\rm i}=\beta_{\rm e}=1$ typical of solar-wind plasma, and a simulation box of size $L_\perp= L_z\approx2.1\times10^{3}$ (we remind the reader that here we use gyrofluid rescaled variables so that in physical units $\tilde L_z\gg \tilde L_\perp$) was discretized with $672^3$ uniformly distributed grid points, corresponding to a (fully dealiased) wavenumber range $0.003\lesssim k_\perp\lesssim 0.67$ (see \citet{CerriAPJ2022} for additional details on the numerical setup). In these simulations, due to successive collisions of these oppositely traveling packets, a quasi-steady turbulent state was reached with different fluctuation properties depending on the large-scale nonlinearity parameter $\chi_0$ (Fig.~\ref{fig:AWcoll_contours}).
By changing the strength of the nonlinear interaction between the AWs, the authors were indeed able to show that dynamic alignment can occur even at weak nonlinearities. Moreover, this scale-dependent alignment combined with a longer lifetime of the turbulent eddies at lower $\chi_0$  facilitates the transition to a tearing-mediated regime. In fact, while at $\chi_0\sim1$ a spectrum $E_{B_\perp}\propto k_\perp^{-3/2}$ clearly emerges (Fig. \ref{fig:AWcoll_spectra}), as well as an associated fluctuations' alignment angle that scales as $\sin\theta_{k_\perp}\propto k_\perp^{-1/4}$ (see Fig. 4 of \citet{CerriAPJ2022}), the transition to reconnection-mediated turbulence is not achieved, being out of reach within the employed resolution. On the other hand, a tearing-mediated regime with a spectrum $E_{B_\perp}\propto k_\perp^{-11/5}$ was clearly observed at $\chi_0<1$ (Fig.~\ref{fig:AWcoll_spectra}; see Fig.4 in \citet{CerriAPJ2022} and associated discussion for the corresponding scale-dependent alignment --- and mis-alignment --- in these regimes), and therefore such regime transition within a weakly nonlinear, dynamically aligned cascade, occurs at even larger scales than those predicted within a critically balanced cascade. Notably, the spectra obtained within these weakly nonlinear regimes exhibited a $-11/5$ range over roughly a decade even at moderately high resolution.

\section{Direct KAW cascade with imbalance: the ion-scale transition range}
\label{sec:cascades}
Alfv\'en wave (AW) turbulence, that plays an important role in magnetized collisionless plasmas such as the solar wind (see \citet{Bruno13, BC16}), is often imbalanced in the sense that the energies ${\mathcal E}^+=\int_0^{+\infty} dk_\perp \int_{-\infty}^{+\infty}  E^+(k_\perp, k_\|) dk_\|$ and ${\mathcal E}^-=\int_0^{+\infty} dk_\perp \int_{-\infty}^{+\infty}  E^-(k_\perp, k_\|) dk_\|$ carried by the waves propagating in the forward and backward directions relatively to the ambient field are unequal \cite{Tu89,Lucek98,Wicks13}. Their ratio $I$ is referred to as the "imbalance parameter" (or, for short, "imbalance"). Here, $E^-(k_\perp, k_\|)$ and $E^+(k_\perp, k_\|)$ is the energy spectra of the corresponding waves. The perpendicular and parallel one-dimensional spectra are defined as $E^\pm_\perp(k_\perp )= \int_{-\infty}^{+\infty}  E^\pm(k_\perp, k_\|) dk_\|$ and $E^\pm_\|(k_\| )= \int_0^{+\infty}  E^\pm(k_\perp, k_\|) dk_\perp$, respectively.

\subsection{Helicity barrier and transition zone: strongly versus moderately nonlinear regimes}\label{sec:simu-imbalance}
While energy is expected to always cascade to small scales, in imbalanced turbulence, the GCH dynamics is the result of conflicting constraints. 
In fact, at ``large'' (fluid) scales, the GCH corresponds, up to a sign, to the usual MHD cross helicity and is expected to develop a direct cascade. At ``small'' (kinetic) scales, on the other hand, the GCH identifies with magnetic helicity, and is instead subject to an inverse cascade \cite{Milo21}. What emerges from these constraints is thus a picture where the GCH displays a ``helicity barrier'' effect \cite{Meyrand21}, where the $k_\perp$-flux of GCH cascading from large MHD scales would not be able to cascade below the ion scales, thus getting ``halted'' before such scales---and then part of this flux may have to be somehow ``redirected'' towards the small parallel scales. 

Simulations of imbalanced turbulence that use an asymptotic form of the two-field  gyrofluid model assuming a sufficiently small value of $\beta_e$ (but nevertheless larger than $\delta^2$ in order for electron inertia to be negligible),
were performed by \citet{Meyrand21}. In these simulations, turbulence  is driven at MHD scales by a negative damping with non-zero cross helicity. Such type of forcing leads to a regime of large imbalance and large nonlinearity parameter that actually breaks the asymptotics underlying the derivation of gyrofluid models. In these simulations, the helicity barrier leads to a strong depletion of the energy flux towards the small perpendicular scales, with a significant energy transfer in the parallel direction, spectrally localized at the ion perpendicular scale. Energy accumulates at large scales and imbalance increases until saturation occurs through strong parallel dissipation. The level of this saturated imbalance depends on the parallel viscosity, thus making it non-universal.
The non-universality associated to a dissipation that depends upon large-scale properties and the fact that {\it ``imbalance produced by large-scale injection of GCH at a prescribed rate is enhanced by wave dispersion"} were pointed out  by \citet{PS19} and \citet{Milo20}.  This dynamics results in a local steepening of the transverse energy and magnetic spectra at a transition  between the MHD and the sub-ion scales, thus qualitatively producing a transition range.

In the above simulations, dissipation results only from hyper-diffusivity terms supplemented in each of the dynamical equations of the model.  In a  kinetic simulation by \citet{Squire22}, performed  with $\beta_i=\beta_e = 0.3$, dissipation is seen to originate from ion-cyclotron resonance (in combination with 4th-order hyper-resistivity, at even smaller scales). The magnetic spectrum again displays an ion transition zone, with  polarity change, characteristic of a situation  where Alfv\'en waves trigger the formation of the ion-cyclotron waves, as observed in the solar wind \cite{Bowen22}.

Differently, in the numerical simulations of the two-field gyrofluid model (taken in the limit of vanishing electron inertia) presented in \citet{PSL22}, 
the driving is ensured by freezing the amplitude of the modes whose transverse wavenumber stands in the first spectral shell and the parallel one corresponds to $k_z (L/2\pi) = \pm 1$ \cite{Siggia78}, a procedure that permits simulations with high imbalance and a nonlinearity parameter that remains moderate, as required by the gyrofluid asymptotics. 
In this setting, differently from the simulations by \citet{Meyrand21}, the injection rates of energy and GCH are not prescribed (being nevertheless approximately constant when the system reaches a stationary regime, at $t \approx 8000 \Omega_i^{-1}$). They in fact just compensate the amount of energy and GCH transferred by the turbulence, from the frozen modes to smaller scales. 
\begin{figure}
    \centering
    \includegraphics[width=0.48\textwidth]{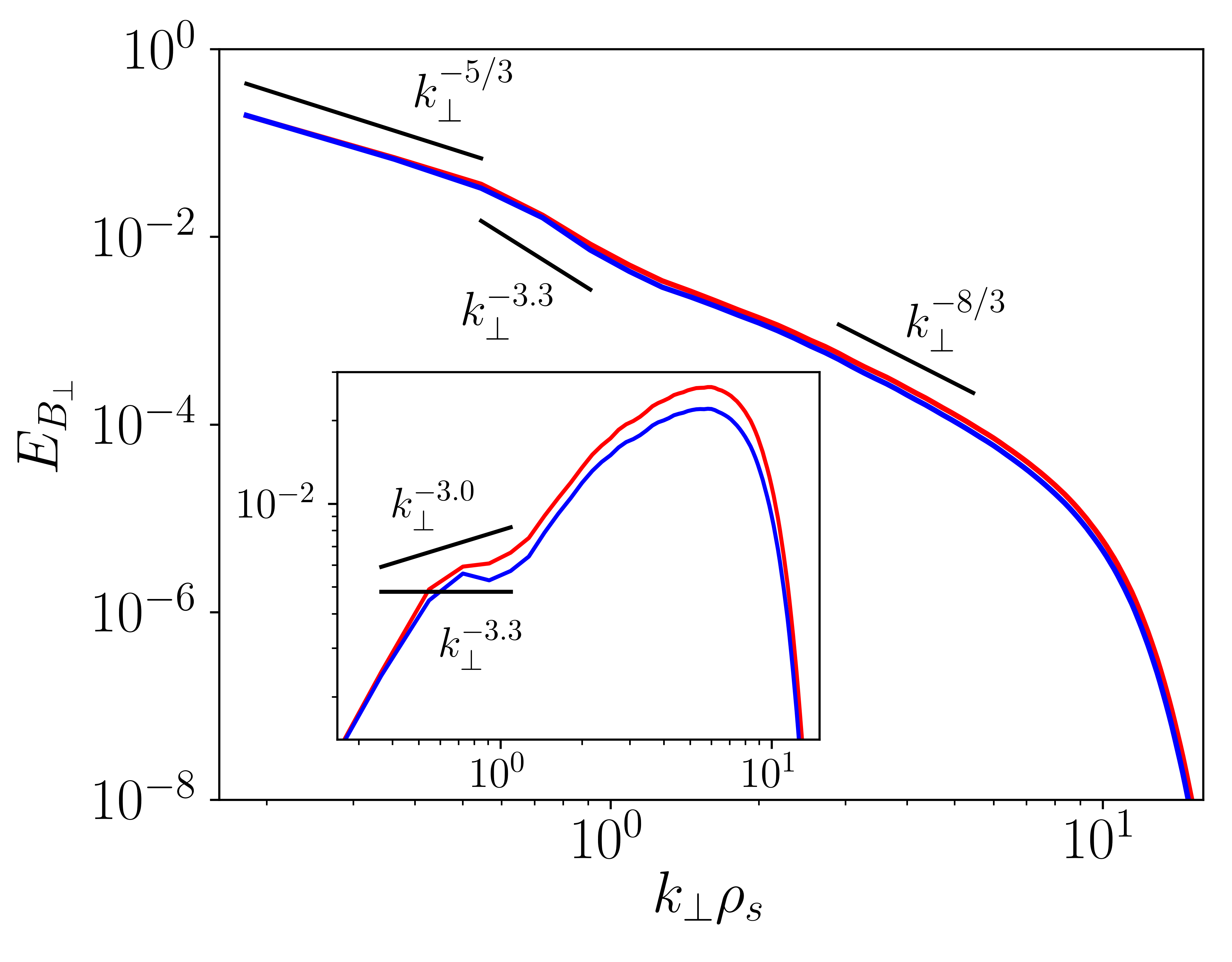}
    \includegraphics[width=0.48\textwidth]{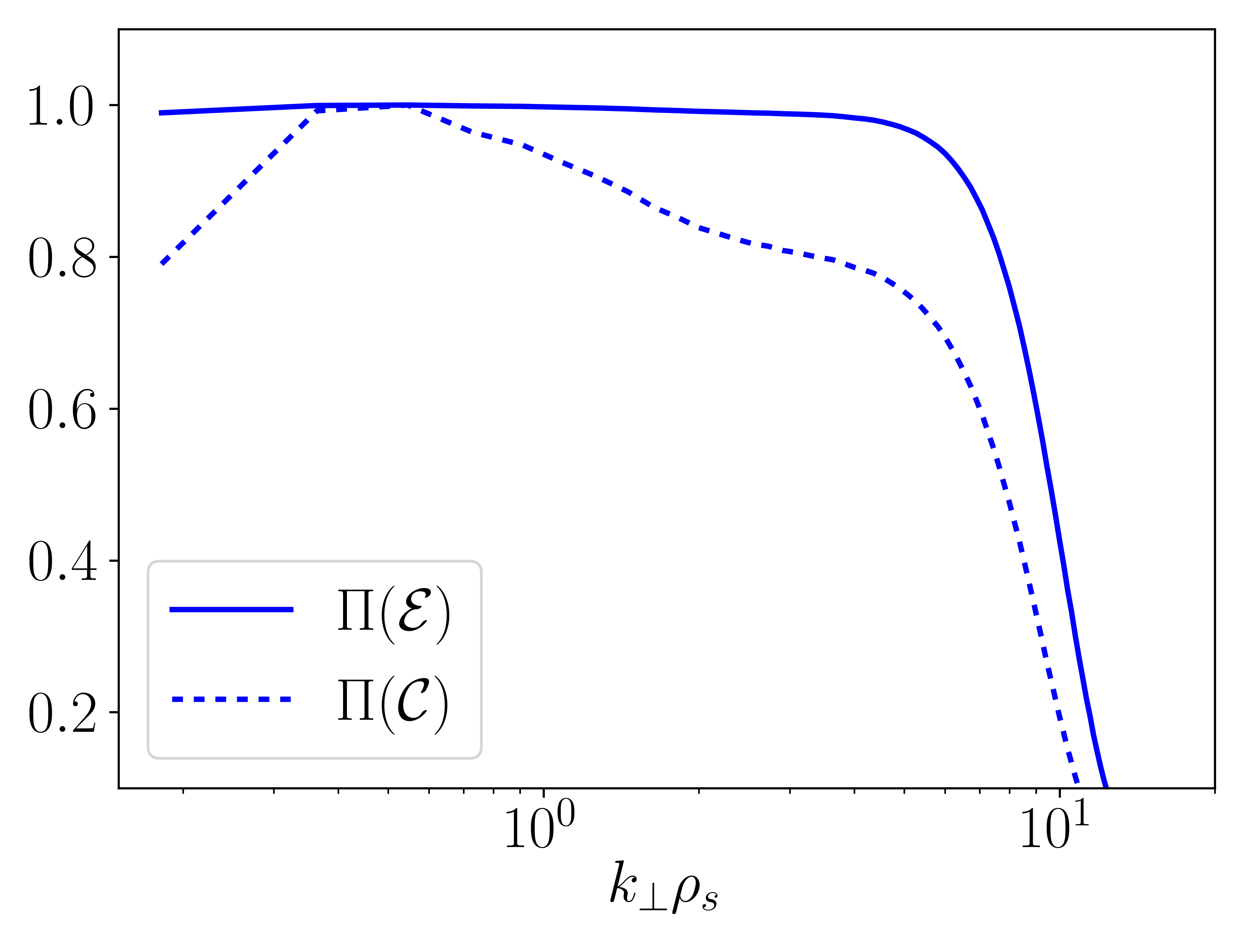}
	\caption{Left: Energy spectra of $B_\perp$ fluctuations versus $k_\perp\rho_{\rm s}$, averaged over the time intervals [8000,14000] (red) and [14000:20000] (blue) (inset: same spectra, compensated by $k_\perp^{3.3}$). Right: perpendicular energy and cross-helicity fluxes, averaged over the time interval [14000:20000], and normalized to their respective maximal averaged values.}
	\label{fig:spectre-imbalance}
\end{figure}

\begin{figure}
	\centerline{
		\includegraphics[width=0.49\textwidth]{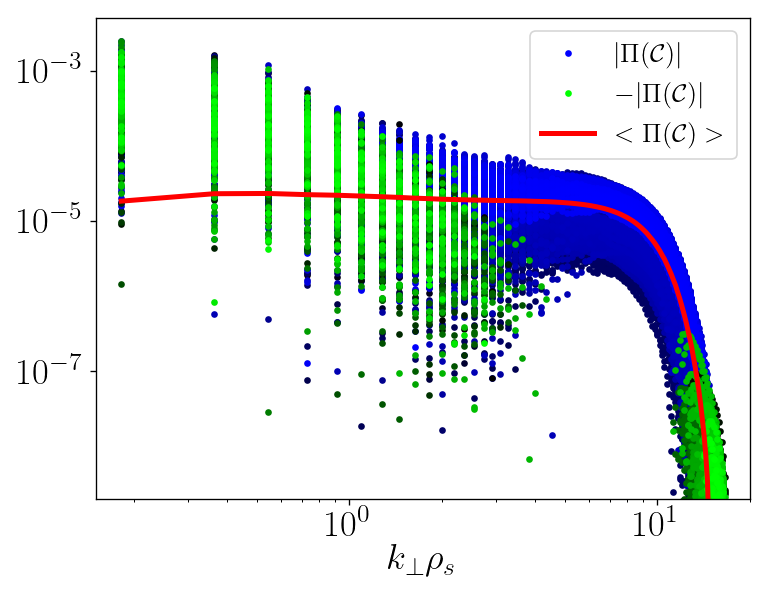}
  \includegraphics[width=0.49\textwidth]{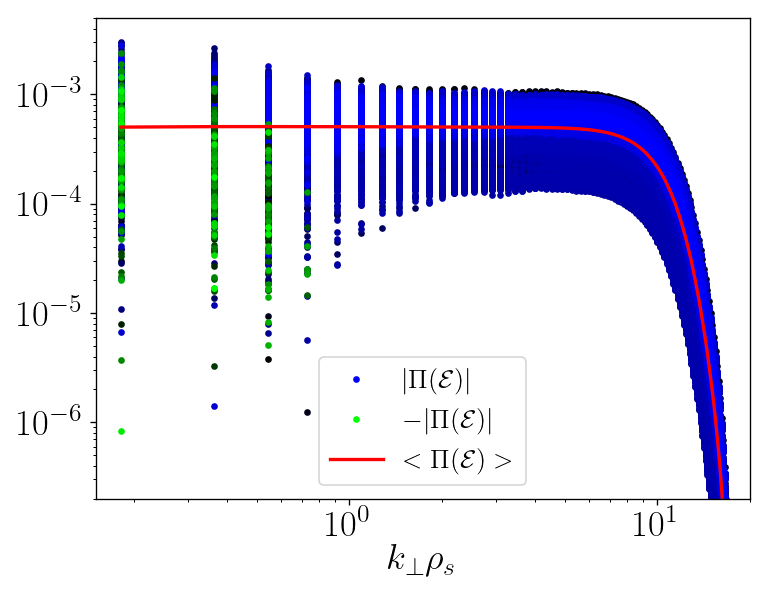}
	}
	\caption{In the conditions of Fig. \ref{fig:spectre-imbalance}, instantaneous GCH (left)  and energy (right)  fluxes when positive (blue) or their opposite when negative (green) in the time interval [14000:20000], together with the averaged fluxes on this interval (red).
}
	\label{fig:Flux-imbalance}
\end{figure}

Figure \ref{fig:spectre-imbalance} (left) displays the energy spectrum of the transverse magnetic field for a simulation performed with $\beta_e = 2$ and $\tau=1$. The  imbalance  is $I= 10$ and the nonlinearity parameter reached in the stationary regime is $\chi = 0.60$. A main observation is that, in this setting also, the transverse energy spectrum of the dominant wave, and consequently, that of the transverse magnetic fluctuations displays a steep transition zone (characterized by a spectral exponent close to $-3.3$) between the Kolmogorov MHD range and the sub-ion range where the spectrum is consistent with the phenomenological prediction $k_\perp^{-8/3}$ of (balanced) KAW turbulence \cite{Boldyrev12}. Furthermore, the spectrum in the transition zone becomes steeper when the imbalance is increased \cite{PSL22}, consistent with Solar Probe observations \cite{Huang_2021}. Interestingly, as seen in Fig. \ref{fig:spectre-imbalance} (right) the time-averaged energy flux remains essentially constant in all the spectral range (with a decrease by no more than $1\%$), while it was reported to decay in \citet{Meyrand21} simulations at large $\chi$. The helicity barrier is, in contrast, conspicuous on the GCH flux which displays a clear decay of about $17\%$ in the transition zone and beyond. 
It is also of interest to display the time fluctuations of the energy and GCH fluxes. Figure \ref{fig:Flux-imbalance} shows that the fluxes can be negative in some spectral zones, especially at large scales, but also at the dissipative scales for GCH, due to an imperfect pinning of the $E^+$ and $E^-$ spectra.
The magnitudes of the fluxes undergo large variations, which are more important for GCH at the end of the transition zone and beyond.

\subsection{Phenomenological modeling}
In \citet{PSL22} a phenomenological nonlinear diffusion model is derived, based on a previous work of \citet{Voitenko16}. In this model, the existence of a transition zone is related to the development of nonlinear interactions between co-propagating waves, which are absent in the MHD range, but becomes possible at sub-ion scales because of the dispersive character of KAWs. This model thus provides an alternative mechanism for the formation of  a transition zone between the MHD range and the sub-ion scales, in regimes where the nonlinear parameter is too small for ion-cyclotron waves to be efficiently excited. It thus seems to be relevant in the simulations presented in \cite{PSL22} which display a transition zone that persists when the parallel dissipation is set to zero. In the following, we briefly review the main elements underlying such mechanism.

We consider a KAW characterized by a wavevector  ${\boldsymbol k}$, undergoing a triadic interaction with two other KAWs with  wavevectors  ${\boldsymbol p}$ and  ${\boldsymbol q}$.  
We are here concerned with the typical time scale of energy transfer, a crucial ingredient in determining the spectral energy distribution. It is possible to estimate this quantity within
the classical turbulence phenomenology whereby two  KAWs with similar wavenumbers $k$ distort each other. As detailed in \citet{PST18}, neglecting the $B_z$ contribution (which is a priori valid for small $\beta_e$) and the  $z$-derivative, the two nonlinear terms of Eq. (\ref{eq:A}), $\nabla_\| \varphi$ and $\nabla_\| N_e$, are respectively associated with two characteristic stretching
frequencies. Assuming a strong turbulent regime involving a critical balance between the characteristic time of the (nonlinear) transverse dynamics and the temporal period of the Alfv\'en waves at the corresponding scale, they are respectively given by $\gamma^{St1}_{k}\simeq (k_\perp^3 B_k)/(M_2 v_{ph})$ and $\gamma^{St2}_{k}\simeq (k_\perp^3 B_k)/(v_{ph})$. Here, the quantity $v_{ph}$ (similar to a phase velocity) is defined as $v_{ph}({\boldsymbol k})=|\omega_{\boldsymbol k}/k_z |$ where $\omega_{\boldsymbol k}$ is the linear frequency of a KAW
of wavevector $\boldsymbol k$ obtained from the linearization of Eqs. (2)-(3). We have used the eigenmode relation $A_{\|,k}=(\beta_e/2) v_{ph}(M_2 \varphi_k/k_\perp^2)$ between the parallel magnetic and the electric potentials in the Fourier space (see Eq. (52) of \citet{PST18}), a relation that appears to be well verified numerically (with a typical accuracy of the order of $10\%$ outside the dissipation range).
The relevant nonlinear frequency is  $\gamma^{St}_k=\rm{max}(\gamma^{St1}_{k},\gamma^{St2}_{k})$, whose wavenumber dependence can be explicited in the three spectral zones defined as the large-scale range (denoted MHD, where $k_\perp\ll k_T$), the weakly dispersive range (denoted WD, where $k_\perp\lesssim k_T$) and the strongly dispersive range (denoted SD, where $k_\perp\gg k_T$). Here $k_T$ refers to a transition wavenumber which is associated with the largest of the ion and sonic Larmor radii. 
While $k_T$ is only weakly sensitive to the level of imbalance, the WD zone extends to larger scales when the imbalance is increased.
Using Table 1 of \citet{PS19} to determine the spectral dependence of $M_2$ and $v_{ph}$, we find that $\gamma^{St}\simeq k_\perp B_k$ in both the MHD and WD ranges, while $\gamma^{St}\simeq k_\perp^2 B_k$ in the SD range.
In this phenomenological evaluation, however, the contributions from co- and counter-propagating waves are not separated and imbalance between the energy of forward and backward propagating waves is not taken into account.

An alternative way of determining the nonlinear frequency is to explicitly write in Fourier space the nonlinear equations for the resonant interaction of a triad with wavevectors ${\boldsymbol k}$, ${\boldsymbol p}$ and ${\boldsymbol q}$. 
We shall use an approach similar (but nevertheless slightly different\footnote{Instead of using the linear decay rate as in \cite{Voitenko11}, which is only valid when two of the three waves have an asymptotically small amplitude, here we estimate the nonlinear transfer rate for three wave interaction in a situation where two waves have finite amplitude.}), to that used by \citet{Voitenko11}.
The 3-wave resonance conditions
simply read ${\boldsymbol k}={\boldsymbol p}+ {\boldsymbol q}$ and 
$\omega_{\boldsymbol k}= \omega_{\boldsymbol p}+ \omega_{\boldsymbol q}$,
where the frequencies are here taken in the limit of vanishing electron inertia. The equations are given in Appendix A of \cite{Milo21}, together with the growth or decay rate for the parametric instability of the ${\boldsymbol k}$-wave (often called pump wave) when its amplitude is finite while those of the ${\boldsymbol p}$ or ${\boldsymbol q}$ waves are much smaller. If two waves (hereafter the ${\boldsymbol k}$ and ${\boldsymbol p}$-waves) are of similar finite amplitude, while the ${\boldsymbol q}$-wave is of smaller amplitude, one can still derive the same equation for the characteristic nonlinear frequency which reads, after correcting a typo in \cite{Milo21},
\begin{eqnarray}
\gamma^2({k_\perp}) &=& \frac{1}{64} \frac{({\widehat {\boldsymbol z}}\bcdot({\boldsymbol p}\times {\boldsymbol q}))^2}{\xi(p_\perp)\xi(q_\perp)}\frac{1}{k_\perp^2 p_\perp^2 q_\perp^2}
\left ( \frac{\sigma_k}{\xi(q_\perp)}-\frac{\sigma_q}{\xi(k_\perp)}\right)
\left(\frac{\sigma_p}{\xi(k_\perp)}-\frac{\sigma_k}{\xi(p_\perp)}\right)
\nonumber\\
&& \nonumber\\
&& \times \left (\sigma_k k_\perp^2 \xi(k_\perp) + \sigma_p p_\perp^2 \xi(p_\perp) + \sigma_q q_\perp^2 \xi(q_\perp)\right )^2 |a_{\boldsymbol k}^{\sigma_k}|^2, \label{eq:decayrate}
\end{eqnarray}
where $\xi =(2/\beta_e)^{1/2}/v_{ph}$, $\sigma_r=\pm 1$ according to the forward or backward direction of propagation of the waves, and where the eigenmode variable $a_{\boldsymbol k}^{\sigma_k}$ satisfies $|a_{\boldsymbol k}^{\sigma_k}|^2 = (8/\beta_e)B_k^2$ \citep{PS19}. Note that, after changing the sign of $\sigma_k$, the same growth rate is recovered by exchanging $p_\perp$ and $q_\perp$ and simultaneously changing $\sigma_p$ and $\sigma_q$ into their opposite, leaving the relative direction of propagation of the waves unchanged. Without loss of generality,  in the following we can thus fix $\sigma_k=+1$.

Equation (\ref{eq:decayrate}) provides an interesting  way to analyze the co-and counter-propagating interactions, as required when there is a large energy imbalance between the waves propagating in opposite directions. We shall assume that the interactions are local, so that the three perpendicular wavenumbers are comparable.
Let us consider the cases where the ${\boldsymbol k}$ and ${\boldsymbol p}$ waves are either co-propagating ($\sigma_k=\sigma_p=1$) or counter-propagating ($\sigma_k=-\sigma_p=1$). Their amplitude can be comparable or, differently, the ${\boldsymbol p}$ wave can have an amplitude much smaller than the ${\boldsymbol k}$ wave.  In both cases, the ${\boldsymbol q}$ wave is chosen to have a small amplitude. We shall address the two cases  $\sigma_q=\pm 1$, using parallel or antiparallel superscript arrows to characterize the nonlinear frequency $\gamma_{R} (k)$ at the wavenumber $k$ in range '$R$', associated with co-propagating or counter-propagating interacting  ${\boldsymbol k}$ and ${\boldsymbol p}$ waves.

\begin{itemize}
    \item{\bf Non-dispersive (MHD) range.} In this range, the phase velocity $v_{ph}$ is constant (and equal to the Alfv\'en speed), $v_{ph}={\rm const.}=v_{\rm A}$.
    
    $[\uparrow\downarrow]$ {\em counter-propagating waves ($\sigma_k=-\sigma_p=1$).} When the waves  ${\boldsymbol k}$ and ${\boldsymbol p}$ are counter-propagating, then ${\sigma_p}/{\xi(k_\perp)}-{\sigma_k}/{\xi(p_\perp)}$ is a constant different from zero. When $\sigma_q=1$, however, the interaction vanishes because ${\sigma_k}/{\xi(q_\perp)}-{\sigma_q}/{\xi(k_\perp)}$ is zero,
    $$\left.\gamma^{\uparrow\downarrow}_{MHD}\right|_{\sigma_q=1} = 0\,,$$
    while it is non-zero when $\sigma_q=-1$, resulting in
    $$\left.\gamma^{\uparrow\downarrow}_{MHD}\right|_{\sigma_q=-1} \simeq k_\perp B_k\,,$$ 
    thus reproducing the RMHD nonlinear frequency obtained above with the simple turbulence phenomenology. 
    Note that, for this choice of parameters ($\sigma_p=\sigma_q=-1$), there is no parametric decay instability in this range since $\gamma(k)$ is purely imaginary.
    
    $[\uparrow\uparrow]$ {\em co-propagating waves ($\sigma_k=\sigma_p=1$).} If the waves denoted by ${\boldsymbol k}$ and ${\boldsymbol p}$ belong to the MHD range, then co-propagating interactions (and associated rate) clearly vanish. In fact, in this range $v_{ph}$ is constant (and equal to the Alfv\'en speed), and thus ${\sigma_p}/{\xi(k_\perp)}-{\sigma_k}/{\xi(p_\perp)}=0$, so 
    $$\gamma^{\uparrow\uparrow}_{MHD}=0\,.$$
    
    \item{\bf Weakly dispersive (WD) range.} In this range, non-negligible corrections to $v_{ph}$ that depend on the wavenumber start to arise, as $v_{ph}(k_\perp)\simeq (1+\alpha k_\perp^2)^{1/2}$, where $\alpha$ is a constant depending on the plasma parameters.
    
    $[\uparrow\downarrow]$ {\em counter-propagating waves ($\sigma_k=-\sigma_p=1$).} If ${\boldsymbol k}$ and ${\boldsymbol p}$ are associated with counter-propagating waves, the term ${\sigma_p}/{\xi(k_\perp)}-{\sigma_k}/{\xi(p_\perp)}$ is of order unity. Then, using the expression for $v_{ph}(k_\perp)$, the term ${\sigma_k}/{\xi(q_\perp)}-{\sigma_q}/{\xi(k_\perp)}$ is $O(k_\perp^2)$ when $\sigma_q=1$, while it is approximately constant when $\sigma_q=-1$. As a result, the interaction rate for $\sigma_q=1$ becomes
    $$ \left.\gamma^{\uparrow\downarrow}_{WD}\right|_{\sigma_q=1}\simeq k_\perp^2 B_k\,, $$
    whereas for $\sigma_q=-1$ it yields
    $$ \left.\gamma^{\uparrow\downarrow}_{WD}\right|_{\sigma_q=-1} \simeq k_\perp B_k\,.$$
    Thus, when $\sigma_p=\sigma_q=-1$ one recovers the rate predicted above with the simple turbulence phenomenology.
    
    $[\uparrow\uparrow]$ {\em co-propagating waves ($\sigma_k=\sigma_p=1$).} When ${\boldsymbol k}$ and ${\boldsymbol p}$ correspond  to  co-propagating waves in the WD range, ${\sigma_p}/{\xi(k_\perp)}-{\sigma_k}/{\xi(p_\perp)}$ scales as $k_\perp^2$. Then, when $\sigma_q=1$, also ${\sigma_k}/{\xi(q_\perp)}-{\sigma_q}/{\xi(k_\perp)}$ scales as $k_\perp^2$ and the corresponding interaction rate is
    $$ \left.\gamma^{\uparrow\uparrow}_{WD}\right|_{\sigma_q=1}\simeq k_\perp^3 B_k\,,$$
    whereas for $\sigma_q=-1$, ${\sigma_k}/{\xi(q_\perp)}-{\sigma_q}/{\xi(k_\perp)}$ is roughly constant and thus
    $$ \left.\gamma^{\uparrow\uparrow}_{WD}\right|_{\sigma_q=-1}\simeq k_\perp^2 B_k\,. $$

    \item{\bf Strongly dispersive (SD) range.} This refers to the small-scale range where the phase velocity scales linearly with the perpendicular wavevector, i.e., $v_{ph}(k_\perp)\simeq k_\perp\gg 1$.

    $[\uparrow\downarrow]$ {\em counter-propagating waves ($\sigma_k=-\sigma_p=1$).} When ${\boldsymbol k}$ and ${\boldsymbol p}$ describe counter-propagating waves in the SD range, their interaction rate is
    $$ \gamma^{\uparrow\downarrow}_{SD}\simeq k_\perp^2 B_k\,, $$
    for both $\sigma_q=1$ or $\sigma_q=-1$.

    $[\uparrow\uparrow]$ {\em co-propagating waves ($\sigma_k=\sigma_p=1$).} Also in the case of co-propagating waves with ${\boldsymbol k}$ and ${\boldsymbol p}$ in the SD range, the result does not depend on $\sigma_q$, and the corresponding interaction rate is
    $$ \gamma^{\uparrow\uparrow}_{SD}\simeq k_\perp^2 B_k\,. $$

\end{itemize}

These scalings have been verified numerically, considering the most unstable triads specified in \citet{Voitenko98a} and using the parameters of the simulation discussed below. 

The present analysis is nevertheless incomplete as, for example, it does not specify the values of the nonlinear frequency, nor the details of the resonance conditions that could contribute to select among the conditions $\sigma_p = \pm \sigma_q$. Some constraints are however to be prescribed in order to  reproduce  classical results.  For consistency with the turbulence phenomenology, it appears that $\sigma_p=\sigma_q$ is a better choice for the MHD range as well as for the WD range for counter-propagating ${\boldsymbol k}$ and ${\boldsymbol p}$ waves. For co-propagating waves, most relevant for the present study, the case $\sigma_p=\sigma_q$ corresponds to a coupling of 3 co-propagating waves which, in the strong-turbulence regime with a large imbalance, can be the most efficient as they all have the largest amplitude.
To summarize, the above discussion leads us to adopt the following scalings:

\begin{itemize}
    \item{\bf counter-propagating waves:}
    \begin{equation}\label{eq:gammaNL_k_counter-prop_summary}
        \gamma^{NL}_k
        \simeq
        \left\{
        \begin{array}{lcr}
         k_\perp B_k\, & \quad & \text{({\it MHD range})}\\
         & & \\
         k_\perp B_k\, & \quad &  \text{({\it WD range})}\\
         & & \\
         k_\perp^2 B_k\, & \quad &  \text{({\it SD range})}
        \end{array}
    \right.
    \end{equation}
    
    \item{\bf co-propagating waves:}
    \begin{equation}\label{eq:gammaNL_k_co-prop_summary}
        \gamma^{NL}_k
        \simeq
        \left\{
        \begin{array}{lcr}
         0\, & \quad & \text{({\it MHD range})}\\
         & & \\
         k_\perp^3 B_k\, & \quad &  \text{({\it WD range})}\\
         & & \\
         k_\perp^2 B_k\, & \quad &  \text{({\it SD range})}
        \end{array}
    \right.
    \end{equation}
    
\end{itemize}

Using the usual Kolmogorov phenomenology (and thus neglecting possible intermittency corrections), it is now possible to predict the magnetic spectra in the various ranges.
We determine the perpendicular magnetic spectrum by writing that the energy flux $\epsilon \simeq \gamma^{Tr}_k B_k^2$ is constant. Here, the energy transfer frequency is given by $\gamma^{Tr}_k =\gamma^{NL}_k$ in the strong turbulence regime or  by $\gamma^{Tr}_k =(\gamma^{NL}_k)^2/\gamma^{Lin}$, where $\gamma^{Lin}=k_z v_{ph}$, in the weak regime and $\gamma^{NL}_k$ refers to one of the nonlinear frequencies determined above.

In the strong turbulence regime, $\gamma^{NL}_k B_k^2\simeq\epsilon\simeq{\rm const.}$ and the above scalings naturally predict a magnetic spectrum $E(k_\perp)\simeq B_k^2/k_\perp \simeq \epsilon^{2/3} k_\perp^{-5/3}$ in the MHD range (neglecting possible effects due to imbalance) and a spectrum $E(k_\perp)\simeq\epsilon^{2/3} k_\perp^{-7/3}$ in the SD range. 

In the weak turbulence regime, the conditions $(\gamma^{NL}_k)^2 B_k^2/(k_z v_{ph})\simeq\epsilon\simeq{\rm const.}$ and $k_z\simeq{\rm const.}$ apply, so the spectrum is $E(k_\perp)\simeq\epsilon^{1/2} k_\perp^{-2}$ in the MHD range (where also $v_{ph}$ is constant) \cite{Galtier00} and it becomes $E(k_\perp)\simeq\epsilon^{1/2} k_\perp^{-5/2}$ in the SD range (where $v_{ph}\simeq k_\perp$) \cite{Galtier03}.

Let us now concentrate on the more delicate WD range. 
Assuming that the dominant interaction is between the co-propagating waves, we predict a magnetic spectrum $E(k_\perp)\simeq\epsilon^{2/3} k_\perp^{-3}$ in the strong regime, while the spectrum steepens to $E(k_\perp)\simeq\epsilon^{1/2} k_\perp^{-4}$ in the weak regime (taking $v_{ph}$ roughly constant). For counter-propagating waves, the slopes are identical to those in the MHD range. 

Before proceeding further in the discussion, we summarize here the predicted scaling in the weak and strong regimes for cases dominated by counter- or co-propagating interactions:
\begin{itemize}
    \item{\bf Strong turbulence:}

    1) {\em counter-propagating-dominated interactions}
    
    \begin{equation}\label{eq:B-spectrum_strong_counter-prop-dominated_summary}
        E(k_\perp)
        \sim
        \left\{
        \begin{array}{lcr}
         \epsilon^{2/3} k_\perp^{-5/3}\, & \quad & \text{({\it MHD range})}\\
         & & \\
         \epsilon^{2/3} k_\perp^{-5/3}\, & \quad &  \text{({\it WD range})}\\
         & & \\
         \epsilon^{2/3} k_\perp^{-7/3}\, & \quad &  \text{({\it SD range})}
        \end{array}
    \right.
    \end{equation}

    2) {\em co-propagating-dominated interactions}

    \begin{equation}\label{eq:B-spectrum_strong_co-prop-dominated_summary}
        E(k_\perp)
        \sim
        \left\{
        \begin{array}{lcr}
         \text{N/A}\, & \quad & \text{({\it MHD range})}\\
         & & \\
         \epsilon^{2/3} k_\perp^{-3}\, & \quad &  \text{({\it WD range})}\\
         & & \\
         \epsilon^{2/3} k_\perp^{-7/3}\, & \quad &  \text{({\it SD range})}
        \end{array}
    \right.
    \end{equation}

  \item{\bf Weak turbulence:}

    1) {\em counter-propagating-dominated interactions}
    
    \begin{equation}\label{eq:B-spectrum_weak_counter-prop-dominated_summary}
        E(k_\perp)
        \sim
        \left\{
        \begin{array}{lcr}
         \epsilon^{1/2} k_\perp^{-2}\, & \quad & \text{({\it MHD range})}\\
         & & \\
         \epsilon^{1/2} k_\perp^{-2}\, & \quad &  \text{({\it WD range})}\\
         & & \\
         \epsilon^{1/2} k_\perp^{-5/2}\, & \quad &  \text{({\it SD range})}
        \end{array}
    \right.
    \end{equation}

    2) {\em co-propagating-dominated interactions}
    
    \begin{equation}\label{eq:B-spectrum_weak_co-prop-dominated_weak_summary}
        E(k_\perp)
        \sim
        \left\{
        \begin{array}{lcr}
         \text{N/A}\, & \quad & \text{({\it MHD range})}\\
         & & \\
         \epsilon^{1/2} k_\perp^{-4}\, & \quad &  \text{({\it WD range})}\\
         & & \\
         \epsilon^{1/2} k_\perp^{-5/2}\, & \quad &  \text{({\it SD range})}
        \end{array}
    \right.
    \end{equation}
    
\end{itemize}

If the imbalance is large, the dominant wave undergoes a very weak nonlinear interaction with the counter-streaming wave due to the much smaller amplitude of the latter. On the other hand the co-propagating interactions can become dominant in the WD range, especially because the interaction is long lasting as the waves have a small phase velocity difference. We thus expect that in the WD range, a steeper magnetic spectrum will develop, associated with a regime somewhat intermediate between genuinely weak and strong turbulence. In the following, we illustrate this prediction by studying the interaction between two co-propagating plane waves with different wavenumbers.

A model for the time evolution of imbalanced Alfv\'enic turbulence, consistent with the above estimates of the transfer times, is presented in \citet{PSL22}.  It appears as an extension of a previously-developed diffusion model in spectral space \cite{PS19,Milo20}, where the effect of co-propagative waves at sub-ion scales is included by retaining interactions between triad wavenumbers that are comparable but not asymptotically close. The importance of this additional coupling, is expressed as a function of the ratio of the phase velocity $v_{ph}(k_\perp)$ to the Alfv\'en velocity, that tends to zero when approaching the MHD range.  In the quasi-stationary regime, a closed system of equations can be written for $u^\pm = E^\pm(k_\perp)/k_\perp$, where free parameters include the energy transfer rate (taken constant) and the GCH transfer rate assumed to slowly decay in the sub-ion range, in order to model the transfer and dissipation in the parallel direction observed near the ion scale in the direct gyrofluid simulations. Numerical integration of this system clearly shows that the energy spectrum of the most energetic wave (and thus the transverse magnetic energy spectrum) displays a transition zone whose  spectral exponent  is steeper when the strength of  co-propagating waves is increased. On the other hand, it was consistently shown that such transition zone is not present when interactions between co-propagating waves are not retained in the model.

\begin{figure}
    \centering
    \includegraphics[width=0.95\textwidth]{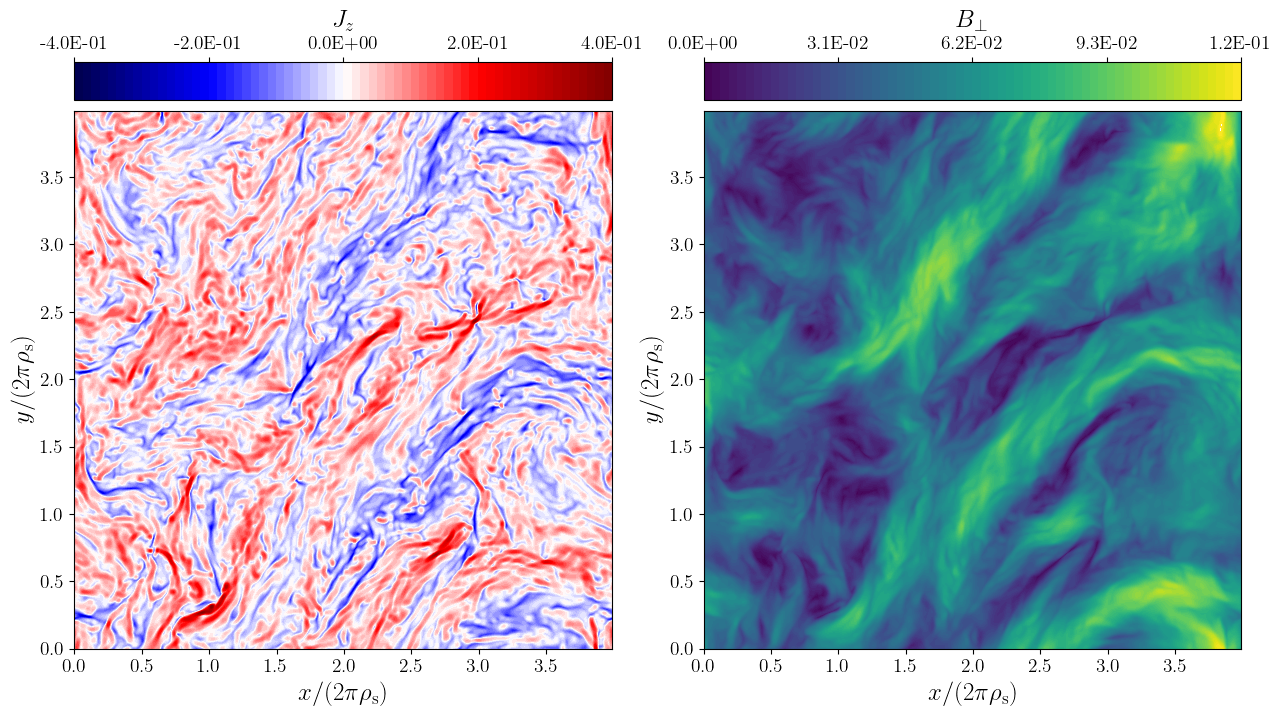}\\
	\includegraphics[width=0.95\textwidth]{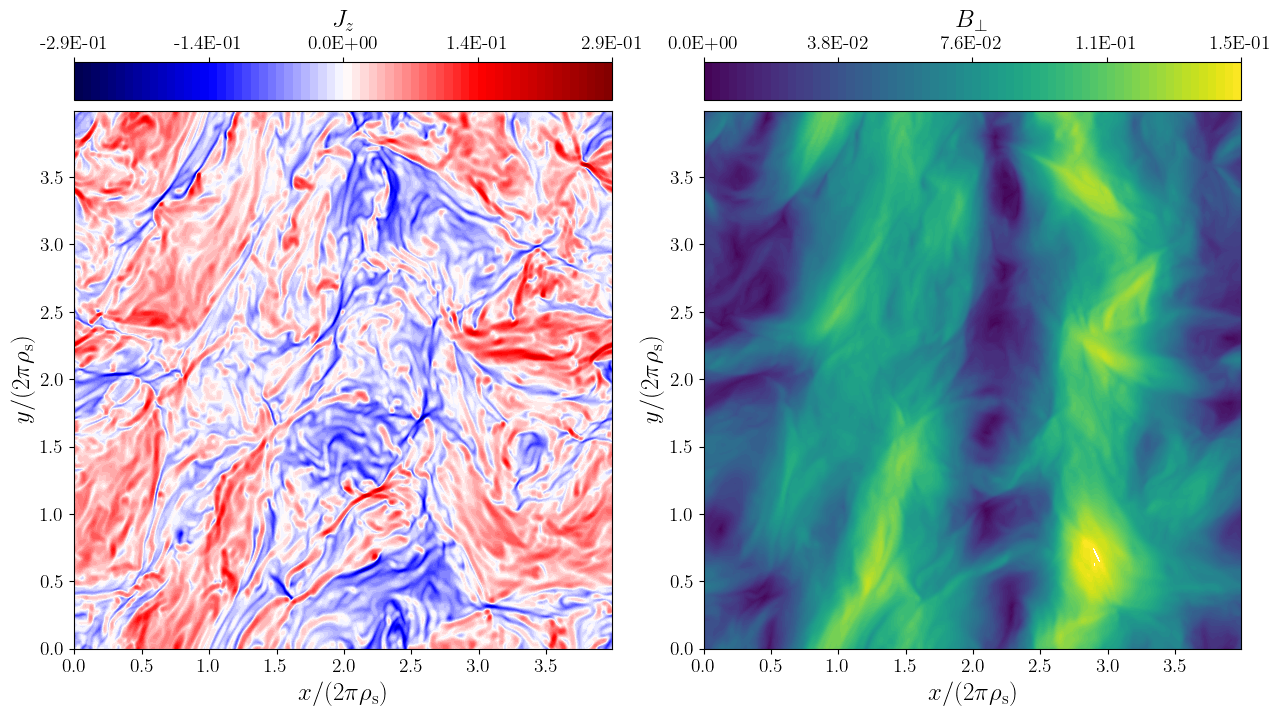}
	\caption{Contours of parallel current density $J_z$ (left) and of perpendicular magnetic field $B_\perp$ (right) in a plane perpendicular to $\boldsymbol{B}_0$ generated by the interaction of counter-propagating (top row) and co-propagating (bottom row) kinetic-Alfv\'en plane waves.}
	\label{fig:KAWcoll_contours}
\end{figure}

\subsection{Testing counter- and co-propagating KAW interactions with simulations}

\begin{figure}
    \centering
    \includegraphics[width=0.9\textwidth]{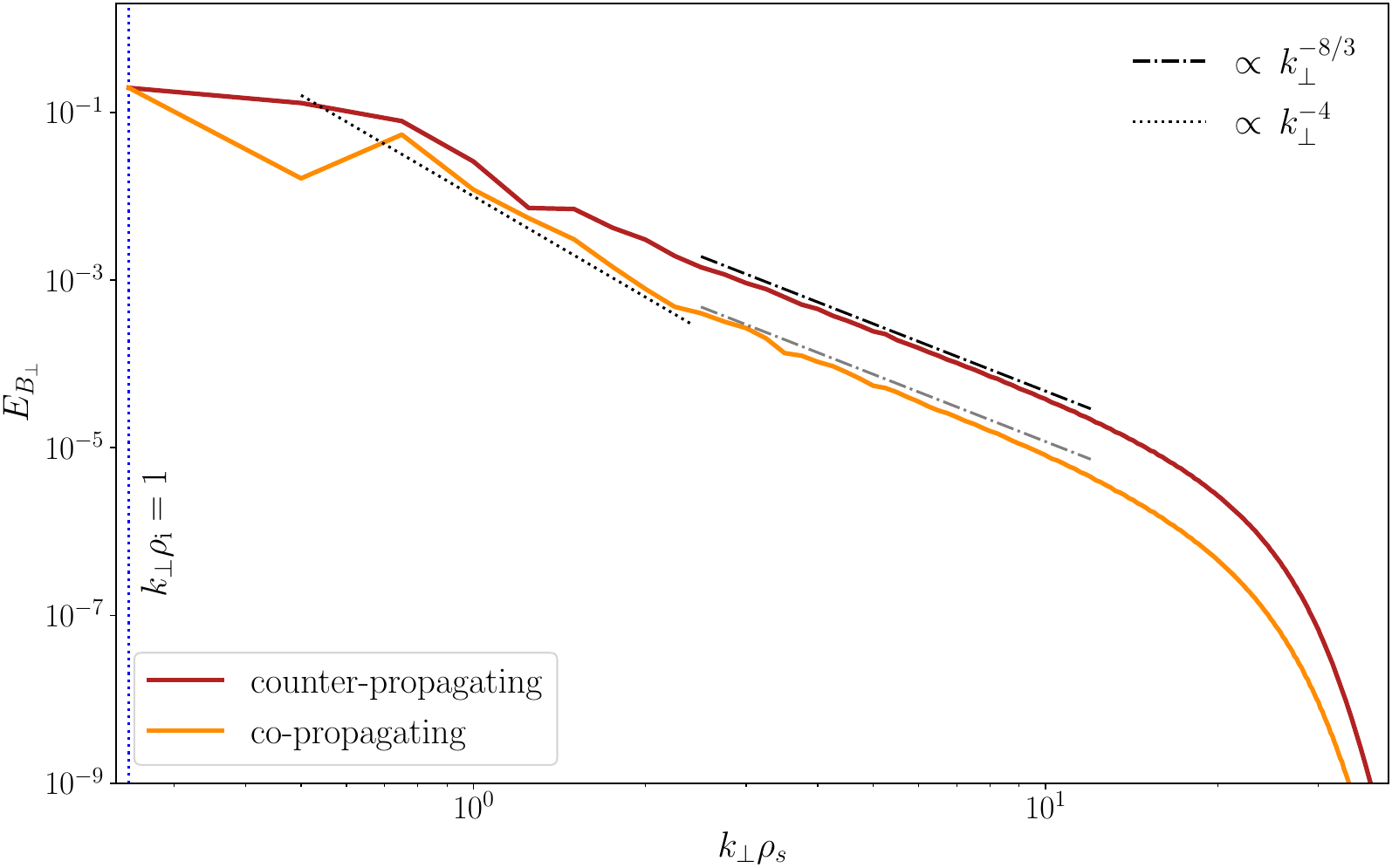}
	\caption{Energy spectra of $B_\perp$ fluctuations versus $k_\perp\rho_{\rm s}$ due to the interaction of counter-propagating (red) and co-propagating (orange) kinetic-Alfv\'en waves. Spectra are averaged over a quasi-steady turbulent state, and relevant power laws are shown as a reference.}
	\label{fig:KAWcoll_spectra}
\end{figure}

In order to validate the influence of the interaction between co-propagating waves on the existence of a transition zone, we performed numerical simulations of decaying turbulence with $\beta_i = 1/2$ and $\beta_e = 1/16$ (so $\tau= 8$), when the initial conditions are two kinetic Alfv\'en plane waves, either counter- or co-propagating, with wavevectors given by 
${\boldsymbol k}^- = (\frac{1}{4},0,\frac{1}{4})$, ${\boldsymbol k}^+  = (0,\frac{1}{2},-\frac{1}{2})$ and by ${\boldsymbol k}^- = (\frac{1}{4},0,\frac{1}{4})$, ${\boldsymbol k}^+  = (0,\frac{3}{4},\frac{1}{2})$, respectively
(we recall to the reader that the background field $\boldsymbol{B}_0$ is along $z$). In these simulations, a box of sizes\footnote{We remind the reader that perpendicular lengths are in $\rho_{\rm s}$ units, while, because of the gyrofluid rescaling, the parallel scales in the physical variables are in fact much larger.} $L_\perp =L_z \approx 25$ has been discretized with $480^3$ uniformly distributed grid points (corresponding to a fully dealiased spectral range $0.25\leq k_\perp \leq 160$), and a combination of Laplacian ($\propto k^2$) and hyper-diffusion ($\propto k^8$) has been employed to properly dissipate energy at the smallest scales.
As expected, turbulence is more developed in the case of counter-propagating waves (Fig.~\ref{fig:KAWcoll_contours}), where the rms (root mean square) current reaches a larger maximum and the energy decay is much more important (not shown).
Both simulations display a $-8/3$ spectrum deep in the sub-ion range (steeper than the $-7/3$ spectrum, possibly because of intermittency effects \cite{Boldyrev12}), with a transition zone with a spectral exponent close to  $-4$ clearly visible in the case of co-propagating wave (Fig.~\ref{fig:KAWcoll_spectra}). Inspection of the phase velocity $v_{ph}$ calculated with the parameters of the simulation shows that the transition above which $v_{ph}\sim k_\perp$  occurs smoothly, and a clear linear scaling is actually achieved with a good level of accuracy (with a margin of $1.3\%$) only at a transition wavenumber $k_\perp \gtrsim 2$. This explains that the transition zone extends to scales as small as $\rho_s/2$, much smaller than $\rho_i$.

These results show that  a steep spectral transition range between the MHD cascade and sub-ion-scale turbulence can emerge from a situation where the energy transfer is dominated by nonlinear interactions of co-propagating waves, such as in imbalanced turbulence, especially in situations, like the simulations of Section \ref{sec:simu-imbalance}, where the energy flux is not strongly depleted (thus not leading to an energy accumulation at scales larger than the dispersive range).

\section{ Turbulent regimes in the nonlinear phase of the tearing instability at large and moderate $\tau$}\label{sec:reconnection}

Hamiltonian reduced fluid models are very convenient to address the problem of collisionless magnetic reconnection for which dissipative effects, necessary for well-posedness in the turbulent regime, should be located at the smallest resolved scales.
Previous investigations of the cold-ion regime, using a reduced description that appears as a limit of our two-field model, have shown that 2D collisionless magnetic reconnection can trigger fluid-like secondary instabilities \cite{Del03, Del05,Del06, Del11, Gra07,Gra09}. At low-$\beta_e$, Kelvin-Helmholtz (KH) or Rayleigh-Taylor-like instabilities (the preeminence of one or the other depending on the value of $\rho_s/d_e$), can lead to the development of turbulence. In this Section, we extend this investigation to moderate and large values of the parameter $\tau$ and analyze the influence of this temperature ratio on  the nonlinear evolution of magnetic islands,  as well as on the properties of the turbulence driven by  the secondary instabilities \cite{GTLPS23}.
We consider two cases representative of the finite-$\tau$ and large-$\tau$ regimes, namely $\tau=1$  and $\tau=100$, and use, for the simulations (performed with the normalization of  Eqs. (\ref{contfin})-(\ref{qnfin}) and $\hat{\rho_s}/L = 1$), the following parameters:
\begin{equation}
    \frac{d_e}{\rho_s} = 0.223, \quad \beta_e = 0.02, \quad \frac{d_i}{\rho_s} = 10, \quad \delta^2 = m_e/m_i = 5 \times 10^{-4}. \label{para1}
\end{equation}
We consider an equilibrium current sheet given by
\beq \label{equilgf}
\varphi^{(0)} = 0, \quad \quad \apar^{(0)} = A_{\parallel 0}^{eq} /\cosh^2 \left( x \right), \quad \quad \bpar^{(0)}=0,
\eeq
with $A_{\parallel 0}^{eq}=1.299$ to ensure that $\mathrm{max}_x(B_{y}^{eq}(x))=1$, where $B_{y}^{eq} (x)=-d \apar^{(0)}/dx$ is the amplitude of the equilibrium magnetic field. 
The tearing stability parameter \cite{Fur63} for this equilibrium is given by 
\beq
\Delta' = 2 \frac{\left(5- k_y^2\right) \left( k_y^2+3\right)}{ k_y^2 ( k_y^2+4)^{1/2}}.
\eeq
The equilibrium (\ref{equilgf}) is initially tearing unstable when $\Delta ' >0$, which corresponds to wave numbers $k_y = \pi m/L_y < \sqrt{5}$.

\begin{figure}
	\centering
		\includegraphics[width=0.8\textwidth]{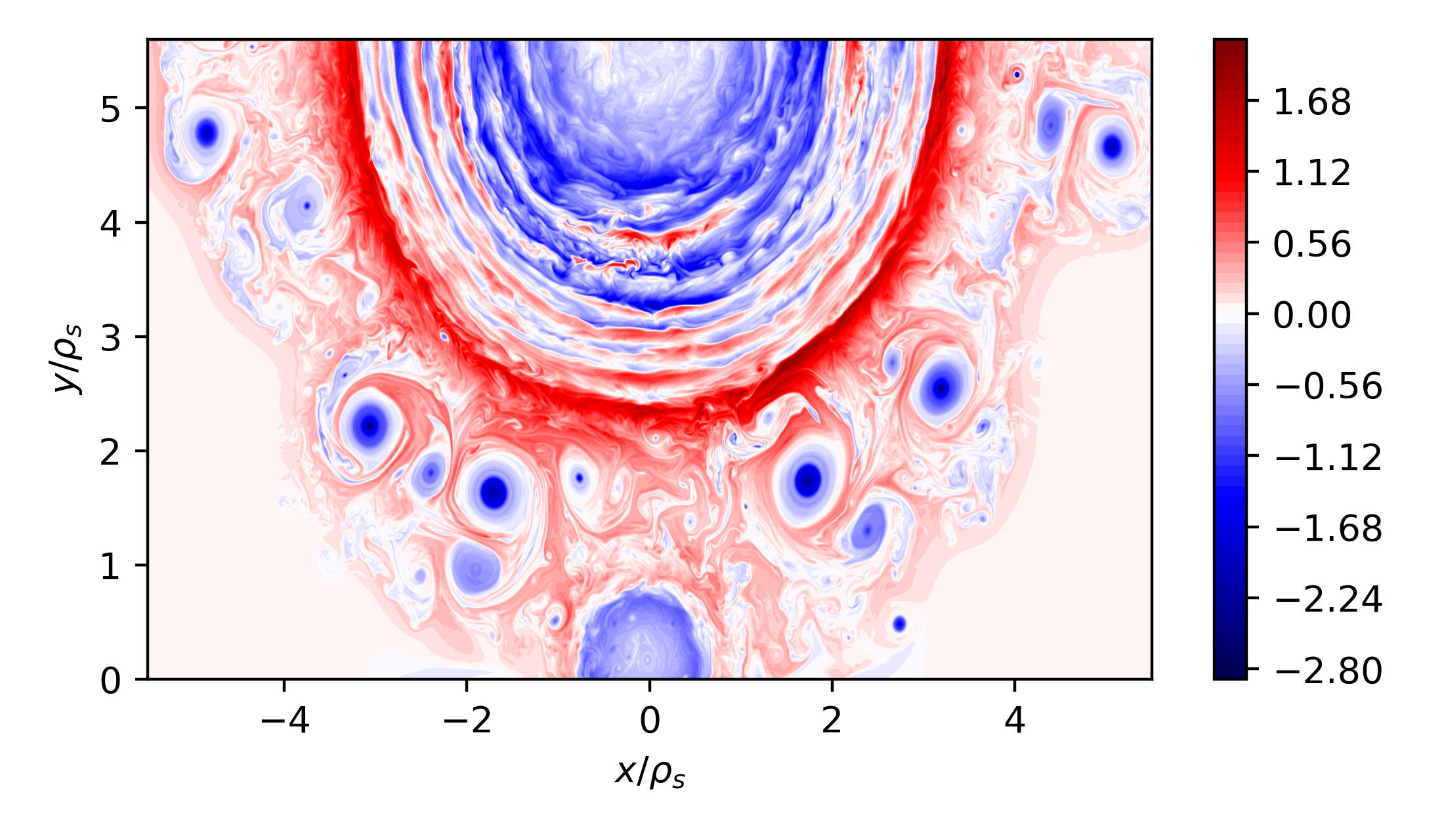}\\
		\includegraphics[width=0.8\textwidth]{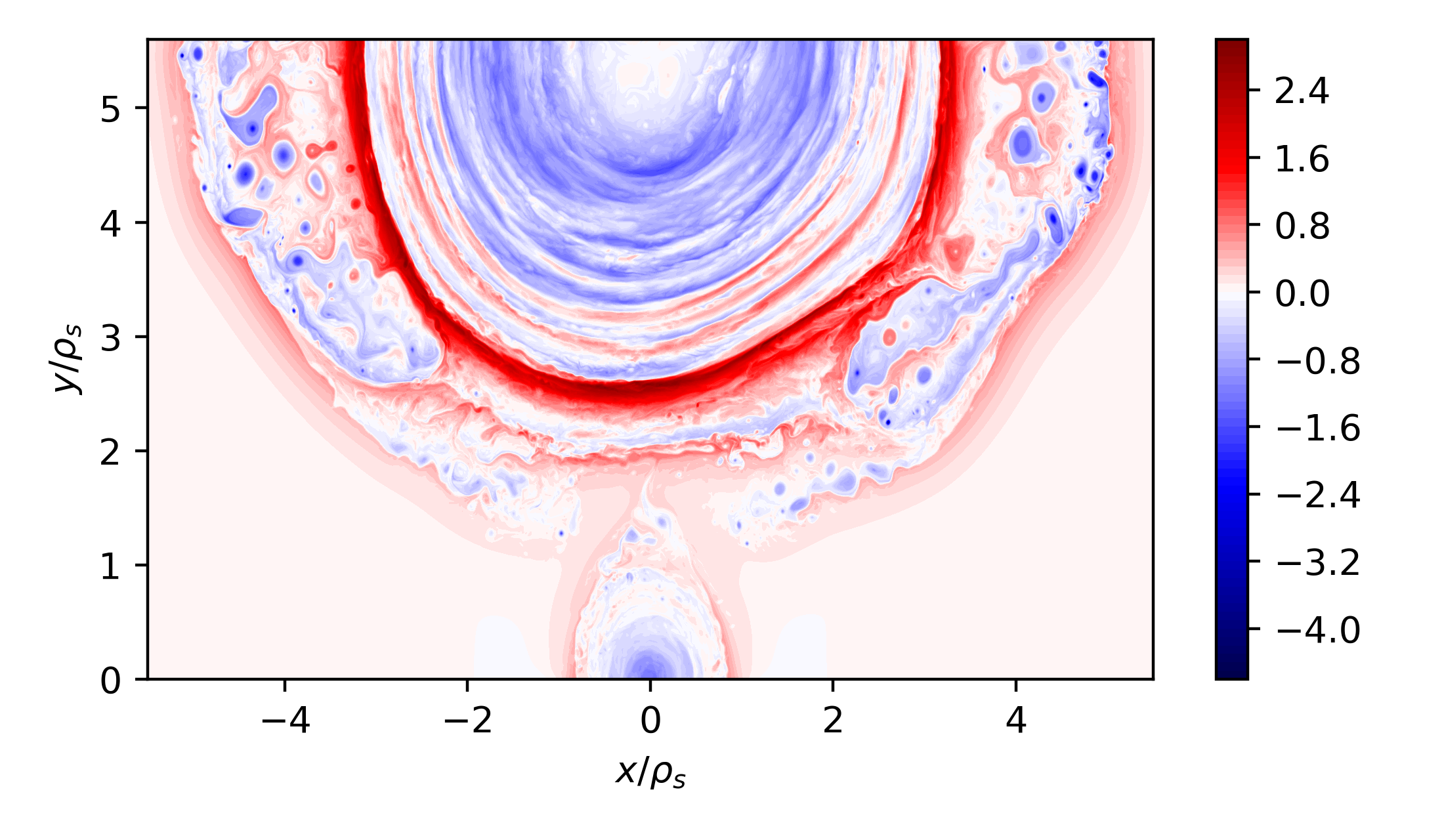}
	\caption{Color-scale plot of the out-of-plane electron gyrocenter velocity $U_e=\lapp \apar$, showing the evolution of the turbulence outside the separatrix for $\tau=100$ (top) and $\tau=1$ (bottom), in the upper-half  simulation domain.}
	\label{fig:Rec-sec}
\end{figure}

We use a grid of $2080^2$ points within a 2D domain defined as $-2.2 \pi \leq x \leq 2.2\pi$ and $-1.8\pi \leq y \leq 1.8\pi$. To introduce numerical dissipation, we apply filters as described in \cite{Lel92}. These filters are designed to damp fluctuations at scales for which $k_{\perp} > 500$, in such a way that the dissipation acts at scales much smaller than the electron Larmor radius.  

In both the cases $\tau=1$ and $\tau=100$, simulations reveal the existence of strong electronic velocity shears, aligned with the separatrices, that initiate KH instabilities. These instabilities lead to the generation of turbulence and to the formation of eddies in a region outside the magnetic island.

In the case $\tau = 100$, the linear growth rate of the tearing mode is larger than for the case $\tau = 1$.
In the early fast growth phase, the outflow directed towards the interior of the island generates a mushroom-shaped structure, symptomatic of a Rayleigh-Taylor instability.  Electron jets collide and generate turbulence at the center of the island, as also reported in the small-$\tau$ regime \cite{Del03, Del06, Gra07, Gra09}.
The strong shear taking place at the separatrices, leads to stretching and distortion of the magnetic field lines, triggering reconnection and then the formation of magnetic eddies. 

The top panel of Fig. \ref{fig:Rec-sec} displays, for $\tau=100$, the out-of-plane electron gyrocenter velocity $U_e=\lapp \apar$, with a number of vortex structures on a broad range of scales that have propagated outside the island and whose size is significantly larger than $\rho_e$ (as shown in \cite{GTLPS23}, the largest structures can be fitted with a Gaussian profile whose FWHM is about $1.3\ d_e \approx9.2\,\rho_e$), indicating that their formation mechanism is insensitive to the dynamics at the electron Larmor scale. Similar vortices are also observed in EMHD simulations, where electron FLR effects are absent. The precise shape of EMHD vortices is however different, with the presence of an inner core whose width appears to be influenced by dissipative effects \cite{GTLPS23}. 
The spectrum of magnetic fluctuations developed in the present simulations has a power law shape at sub-$d_e$ scales, with a spectral index of $-4$ (not shown). This observation is not consistent with phenomenological predictions in the strong turbulence regime (where a $-11/3$ range is expected \cite{PST18,Meyrand2010}), possibly reflecting an effect of the central island and current-sheet structures. Further simulations in the context of homogeneous and isotropic turbulence are thus necessary to address this issue.

In contrast, simulations with $\tau=1$, which show a similar development of the turbulent current layer at the edge of the islands, lead to the formation of smaller vortices (Fig. \ref{fig:Rec-sec}, bottom panel). 
Their size being closer to $\rho_e$, the question arises whether their formation could be associated with the presence of electron FLR effects. To investigate this issue, a simulation where the FLR term in Eq. (\ref{eq:gyro-Ne-phi}) was suppressed, has been performed. Interestingly, this simulation required a more extended filter, indicating that FLR terms have a regularizing effect. In this simulation, we note a significant reduction of magnetic vortices and an absence of current structures of sizes similar to $d_e$ or smaller. This point is discussed in more details in Section \ref{sec:limitation} in the context of homogeneous turbulence.

\section{Limitations of the two-field model}\label{sec:limitation}
In Section \ref{sec:reconnection}, it was mentioned that, when $\tau=1$, the formation of vortices resulting from the KH instability triggered by strong electron velocity shear at the edge of the magnetic island, only occurs in the presence of the (subdominant) electron FLR-correction terms in the two-field model. The size of these vortices is significantly smaller than $d_e$ and in fact approaches $\rho_e$. Since the two-field model does not systematically take into account all the electron FLR effects, the dynamics that is observed at scales approaching the electron Larmor radius in these simulations is not fully trustworthy. 
In order to better understand the limitations of the model, simulations of homogeneous turbulence in  domains of various sizes have been performed, with a number of collocation points equal to $4000^2$. These simulations have been carried out with the model formulated in the non-dimensional variables used in Eqs. (\ref{eq:Ne})-(\ref{eq:gyro-Ne-phi}).

\begin{figure}
    \centering
    \includegraphics[width=\textwidth]{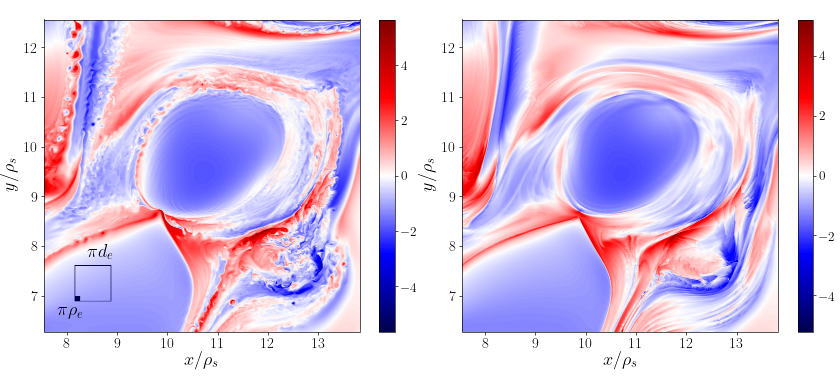}
	\caption{Color-scale plots of $\lapp \apar$ from a simulation of decaying turbulence with (left) or without (right) the electron-FLR correction, for $\beta_e=0.02$, $\tau=1$. The simulations are initialized with homogeneous random fluctuations. Only a quarter of the whole computational box [0, $8 \pi \rho_s$] is shown, at a time ($t=1.375 \Omega_i^{-1}$) taken within the early period of turbulence decay.  On the left panel, squares with sides $\pi d_e$ and $\pi \rho_e$ are superimposed.}
	\label{fig:turb-tau=1-smallbox}
\end{figure}

In Fig. \ref{fig:turb-tau=1-smallbox}, we display $\lapp \apar$ at the same instant of time in a sub-domain of two simulations when including  (left) or not (right)  the electron FLR term. The simulations have been initialized with random fluctuations, taking the same parameters and a domain of similar size (here $8\pi\rho_s$) as in the $\tau=1$ simulation of Section \ref{sec:reconnection}. Similarly to the evolution of the secondary instabilities following the initial reconnection event in Fig. \ref{fig:Rec-sec}, also in Fig. \ref{fig:turb-tau=1-smallbox} small vortices of size comparable to $\rho_e$ are formed in regions of strong shear in the presence of the FLR term (left panel). On the other hand, current sheets display less fragmentation when such FLR term is removed from the gyrofluid equations (right panel of Fig. \ref{fig:turb-tau=1-smallbox}). In fact, increasing the filtering was necessary in order to run the simulation without the FLR term, thus supporting the observation made in the previous Section that FLR terms have a regularizing effect. 
However, the size of the domain used in the simulations of Fig. \ref{fig:turb-tau=1-smallbox} limits the aspect ratio of the current sheets, thus preventing the early formation of plasmoids.
This limitation is not present in the case where the same initial conditions are taken at a larger scale, as shown in Fig. \ref{fig:turb-tau=1-largebox} (left panel).

\begin{figure}
    \centering
    \includegraphics[width=\textwidth]{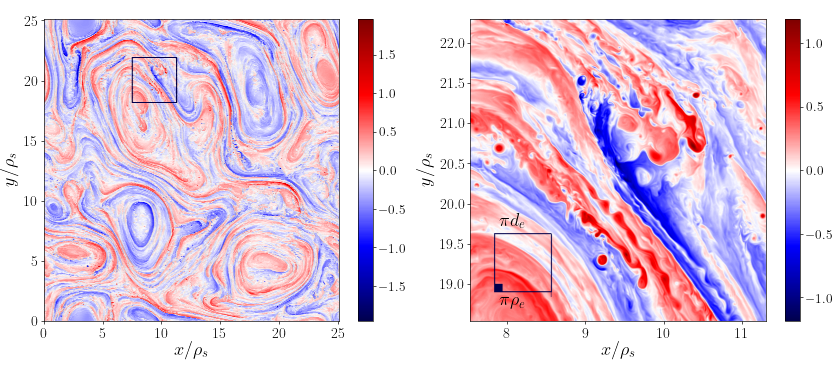}
	\caption{Left: Color-scale plots of $\lapp \apar$ in the whole computational box, from the same simulation of Fig. \ref{fig:turb-tau=1-smallbox} at a later time $t=6$. Right: zoom of the region enclosed in the square displayed in the left panel, along with the characteristic sizes $\pi d_e$, $\pi \rho_e$ as in fig. \ref{fig:turb-tau=1-smallbox} (left).}
	\label{fig:turb-tau=1-smallbox_late}
\end{figure}

\begin{figure}
    \centering
    \includegraphics[width=0.9\textwidth]{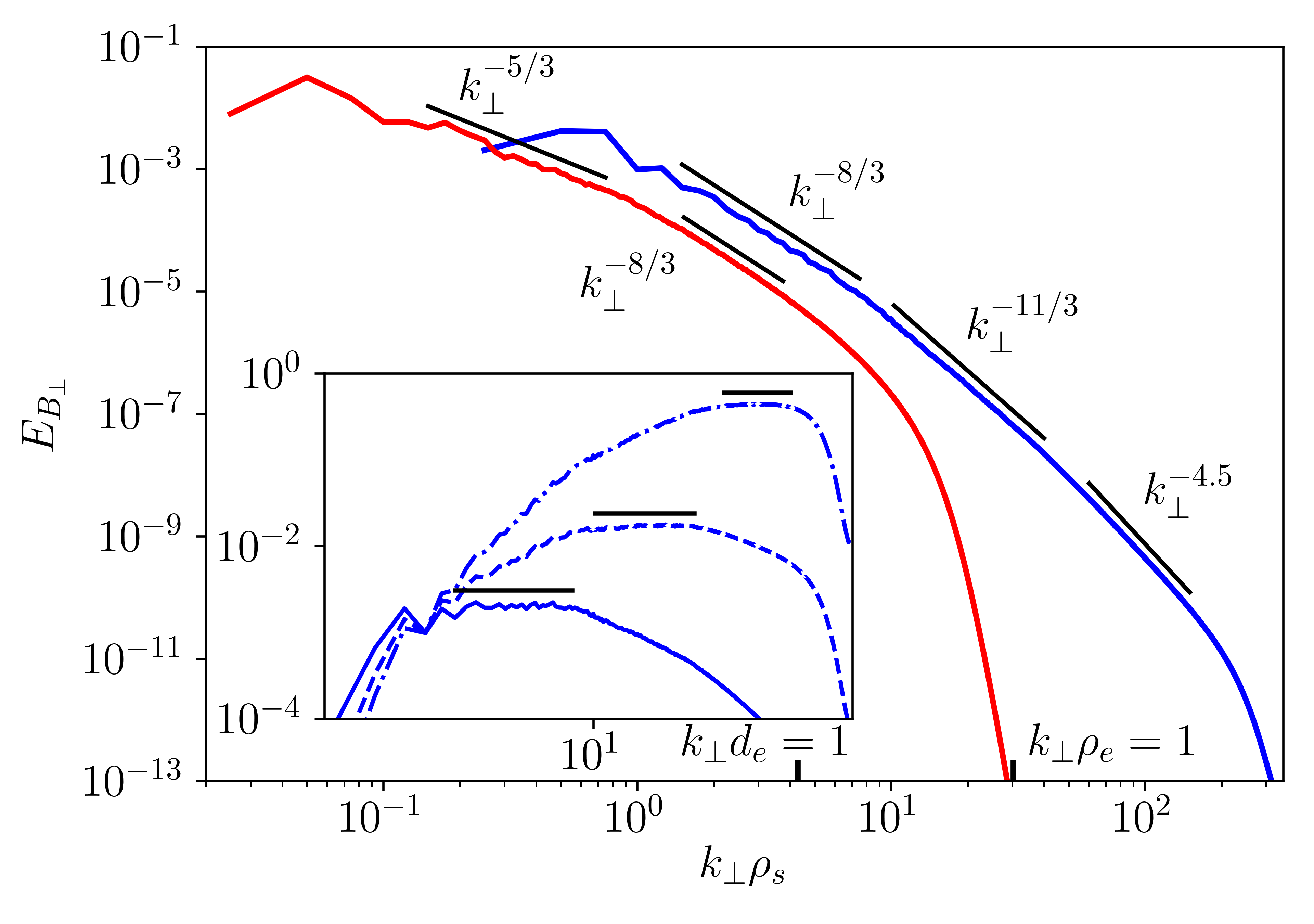}
	\caption{Energy spectra of $B_\perp$ fluctuations versus $k_\perp\rho_{\rm s}$. Blue: same run as in fig. \ref{fig:turb-tau=1-smallbox}, averaged between $t=5$ and $t=7$. Red: same run as in fig. \ref{fig:turb-tau=1-largebox}, averaged between $t=20$ and $t=30$. Inset: blue spectrum, compensated by $k^{8/3}$ (solid line), $k^{11/3}$ (dashed line), $k^{4.5}$ (dash-dotted line)}
	\label{fig:small-large-Bspectra}
\end{figure}

\begin{figure}
    \centering
    \includegraphics[width=\textwidth]{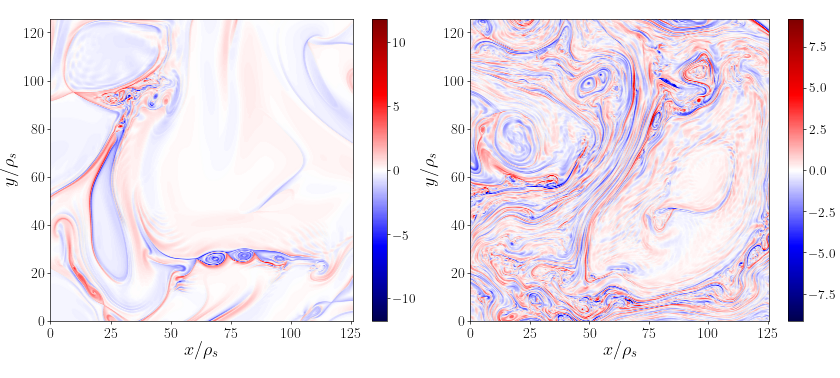}
	\caption{For a decaying run in a box [0, $80 \pi \rho_s$] with random initial fluctuations and $\beta_e=0.02$, $\tau=1$, color-scale plots of the out-of-plane electron gyrocenter velocity $U_e$ in a half of the computational box, at times $t=10 \Omega_i^{-1}$ (left) and $t=24 \Omega_i^{-1}$.}
	\label{fig:turb-tau=1-largebox}
\end{figure}

Figure \ref{fig:turb-tau=1-smallbox_late} displays a later time of the simulation shown in Fig. \ref{fig:turb-tau=1-smallbox}, when using the full gyrofluid model (i.e., including the FLR term), the right panel  being a zoom of the full box (left panel). The global view on the left panel of Fig. \ref{fig:turb-tau=1-smallbox_late} displays multiple current sheets brought together by  the motion of large-scale coherent structures around which some of them are rolling up in a spiral shape structure. These current sheets are thickened by competing KH and reconnection instabilities inside the layer, leading to the formation of many vortices at the electron Larmor radius scale. In Fourier space, the magnetic energy spectrum, displayed in Fig. \ref{fig:small-large-Bspectra}, shows three power-laws associated with the sub-ion range (spectral index close to $-8/3$, obtained by \citet{Boldyrev12} when taking into account monofractal intermittency corrections), the sub-$d_e$ range (spectral index close to $-11/3$ predicted in \cite{Meyrand2010}) and a sub-$\rho_e$ range (spectral index of $-4.5$, consistent with Cluster observations in the terrestrial magnetosheath \cite{Huang14})). These findings, which show consistency with space plasma measurements, deserve confirmation using a model which includes a comprehensive description of electron Larmor radius corrections.

We now display in Fig. \ref{fig:turb-tau=1-largebox} two snapshots of a simulation using the full model with the same physical parameters as above, except that the size of the domain is now an order of magnitude larger, i.e., $80\pi\rho_s$. In this case, scales comparable to the electron Larmor radius are inside the dissipation zone. We note a qualitative similarity with the simulations in the cold-ion regime reported in \cite{Borgogno22}. The most obvious difference with the previous smaller-box simulation (left panel of Fig. \ref{fig:turb-tau=1-smallbox}) is that we now observe vortices at all the scales (including meso-scale plasmoid chains), resulting from the turbulent dynamics. 
In addition to the current sheets, two types of structures are clearly visible at early times (left panel), namely plasmoids formed in extended current sheets as a result of reconnection (lower part of the left panel of Fig. \ref{fig:turb-tau=1-largebox}), and smaller vortices produced by the KH instability that develops in the outflow of a reconnection site (top left part of the left panel of Fig. \ref{fig:turb-tau=1-largebox}). In the late phase of the simulation (right panel), the structures undergo nonlinear interactions, leading to a fully developed turbulent regime where it is difficult to distinguish between structures originated by the two different processes. The associated time-averaged magnetic energy spectrum, displayed in Fig. \ref{fig:small-large-Bspectra}, shows a $-5/3$ MHD range followed by a $-8/3$ sub-ion range, limited at small scale by dissipation. Note that this simulation, which is performed in the spectral range of strict validity of the model, does not resolve the sub-electron scales (and especially $\rho_e$). The question arises whether the active dynamics observed at these scales in the previous simulation, could affect the evolution of meso-scale turbulence.

In order to address a situation where dissipation acts at scales smaller than the electron Larmor radius, and also to extend the validity of the gyrofluid model to larger values of the $\beta_e$ parameter, there is a need for a more sophisticated model taking into account both the ion and electron FLR effects (and also possibly the Landau damping terms, as shortly discussed below). 
It is also necessary to include the coupling to the full ion motion, with dynamical equations for the ion gyrocenter density and parallel velocity. This will also make it possible to capture the coupling of AWs to ion acoustic waves (IAWs), necessary to describe the parametric decay instability at MHD scales (for $\beta_i<1$). Note that a further coupling to electron and ion parallel temperatures would also open the possibility to introduce Landau damping effects (through a Landau-fluid type of closure), particularly important when $\beta_i$  reaches values of order unity \cite{TPS20}.
A 4-field gyrofluid model that takes into account the coupling to the parallel ion velocity and incorporates the full electron FLR effects is detailed in the Appendix. 

\section{Conclusion} \label{sec:conclusion}

The present paper, which includes a review together with new results, illustrates the capability of a relatively simple Hamiltonian two-field model to provide a uniform description, from the MHD to the electron scales, of turbulent space plasmas, when dominated by the dynamics of Alfvén, kinetic Alfvén or inertial kinetic Alfvén waves, and by magnetic reconnection. Such processes are known to constitute the building blocks of heliospheric turbulence. Retaining the whole range of scales within a single  system of equations requires the use of nonlocal operators, most conveniently described in Fourier space, which prescribe constraints on the geometry of the computational domain. This model is thus most suited to address fundamental physical processes, some of which have been reviewed in this paper. We have reported the first numerical evidence of reconnection-mediated MHD turbulence in three dimensional simulations in a regime of moderate and low nonlinearity.  We then analyzed the energy and cross-helicity cascades of imbalanced Alfv\'enic turbulence and in particular the existence of a  transition zone at ion scales that, at moderate nonlinearity,  appears to be associated with an helicity barrier. In the simulations presented in this paper, where the driving prescribes the energy and imbalance levels at the largest scales, there is no significant drop in the energy transfer rate at the ion scale, in contrast with \cite{Meyrand21}, and the transition zone results from the local dominance of co-propagating interactions in this range. This effect is illustrated by new numerical simulations initialized by co- or counter-propagating plane waves. Simulations of 2D collisionless reconnection at moderate and large values of the ion-electron temperature ratio $\tau$ are also reported, pointing out the onset of turbulence as a result of secondary instabilities that develop at the boundaries of the magnetic island formed by the primary tearing instability of an initially unstable current sheet. Simulations of 2D homogeneous turbulence in same physical conditions are used to point out the role of electron FLR corrections on the generation and the characteristic size of the vortices emerging from the development of a secondary KH instability. The latter observations suggest the need for a model providing a complete description of both ion and electron FLR corrections. A new four-field model, presented in the Appendix, is introduced to meet these requirements. This model, which is also Hamiltonian, does not prescribe conditions on $\beta_e$ and is consistent with the linear theory in regimes where the ion gyrocenters can be viewed as adiabatic and the electrons as isothermal. Perspectives involve simulations of low-frequency turbulence and collisionless reconnection with this new  model in both 2D and 3D frameworks.

\appendix  
\section{Derivation of the two-field gyrofluid model from a four-field gyrofluid model}
In this Appendix, we provide a derivation of the model (\ref{contfin})-(\ref{qnfin}), which is slightly different and more detailed with respect to the one presented in the original reference \cite{PST18}. The starting point for the present derivation indeed differs from that of \cite{PST18} and corresponds to the four-field gyrofluid model

\begin{align}
&\frac{\pa N_e}{\pa t}+ [\gamue \varphi - \rho_s^2 2 \gamde B_z , N_e] - [\gamue \apar , U_e]+\frac{\pa U_e}{\pa z}=0,  \label{4fconte}\\
&\frac{\pa}{\pa t}(\gamue \apar - d_e^2 U_e)+ [\gamue \varphi - \rho_s^2 2 \gamde B_z , \gamue \apar - d_e^2 U_e] +\rho_s^2 [ \gamue \apar , N_e]  \nno \\
& +\frac{\pa }{\pa z}(\gamue \varphi - \rho_s^2 (2 \gamde B_z + N_e))=0,  \label{4fmome}\\
&\frac{\pa N_i}{\pa t}+ [\gamui \varphi + \tau \rho_s^2 2 \gamdi B_z , N_i] - [\gamui \apar , U_i]+\frac{\pa U_i}{\pa z}=0,  \label{4fconti}\\
&\frac{\pa}{\pa t}(\gamui \apar + d_i^2 U_i) + [\gamui \varphi + \tau \rho_s^2 2 \gamdi B_z , \gamui \apar + d_i^2 U_i] -\tau \rho_s^2 [ \gamui \apar , N_i]  \nno \\
& +\frac{\pa }{\pa z}(\gamui \varphi + \rho_s^2 (\tau 2 \gamdi B_z + N_i))=0,  \label{4fmomi}\\
&\gamui N_i -\gamue N_e= \frac{1 - \Gamma_0}{\tau} \frac{\varphi}{\rho_s^2} - (\Gamma_0 - \Gamma_1) B_z -(\gamue^2 -1 ) \frac{\varphi}{\rho_s^2} + \gamue 2 \gamde B_z,  \label{4fqn}\\
&\lapp \apar = \gamue U_e - \gamui U_i,   \label{4famppar}\\
&B_z = -\frac{\beta_e}{2}\left( \tau 2 \gamdi N_i + (\Gamma_0 - \Gamma_1)\frac{\varphi}{\rho_s^2} + 2 \tau (\Gamma_0 - \Gamma_1) B_z \right. \nno \\
&\left. + 2 \gamde N_e - \gamue 2 \gamde \frac{\varphi}{\rho_s^2} + 4 \gamde^2 B_z \right).  \label{4fampperp}
\end{align}
As discussed in Sec. \ref{sec:limitation}, this model has an interest of its own. Indeed, while still being  relatively simpler than gyrokinetic models, it does not assume, unlike the two-field  model (\ref{contfin})-(\ref{qnfin}), that $\beta_e$ and $d_e$ are small parameters. Consequently, it retains the coupling with ion gyrocenter density and parallel velocity dynamics, as well as electron FLR effects. 

Equations (\ref{4fconte})-(\ref{4fmome}) and (\ref{4fconti})-(\ref{4fmomi}) correspond to the continuity and parallel momentum equations for electron and ion gyrocenters, respectively. Equations (\ref{4fqn}), (\ref{4famppar}) and (\ref{4fampperp}) represent quasi-neutrality, parallel and perpendicular components of Amp\`ere's law, respectively.  The model is formulated according to the normalization (\ref{norma}). The additional expressions $N_i$, $U_e$ and $U_i$, corresponding to the ion density and to the electron and ion parallel velocity gyrocenter fluctuations, respectively, are present in Eqs. (\ref{4fconte}) -(\ref{4fampperp}). Their normalization is given by
\begin{align}
&N_i=\frac{L}{\hat{d}_i}\frac{\hat{N}_i}{n_0}, \qquad U_{e,i}=\frac{L}{\hat{d}_i}\frac{\hat{U}_{e,i}}{v_A}.
\end{align}
Also, the gyroaverage operators $\gamue=2\gamde$ and $\gamui=2\gamdi$ correspond, in Fourier space, to operators multiplying Fourier coefficients by $\exp(-(\beta_e/4)d_e^2 k_\perp^2)$ and by $\exp(-\tau \rho_s^2 k_\perp^2/2)$, respectively. We point out that these expressions correspond to those obtained by exact integration in \cite{Bri92}. Alternative expressions, providing a better agreement with linear theory for large values of $(\beta_e/4)d_e^2 k_\perp^2 $ and $(\tau /2) \rho_s^2 k_\perp^2$, could also be used, as discussed, for instance, in \cite{Dor93,Sny01,Man18}.

Unlike the parent gyrofluid model \cite{Bri92} used for the derivation in Ref. \cite{PST18}, the four-field model (\ref{4fconte})-(\ref{4fampperp}) is a closed system.  The gyrofluid model in \cite{Bri92} evolves also parallel and perpendicular gyrocenter pressures and accounts for nonuniform equilibrium magnetic, density and temperature equilibria. Expressions for the heat fluxes and fourth-order moments are not specified, as the closure problem is not addressed.  The four-field model (\ref{4fconte})-(\ref{4fampperp}) can be derived from  $\delta f$ gyrokinetic equations such as those presented in  \cite{Schekochihin09,Kun15}. In particular, the hierarchy of the resulting electron gyrofluid equations is truncated by imposing a quasi-static closure, as described in Ref. \cite{TPS20}. In the present case, in which the hierarchy is truncated at the level of the parallel momentum equation, the quasi-static closure amounts to imposing an isothermal electron fluid. The ion gyrofluid hierarchy, on the other hand, is truncated using the closure described in  \cite{Sco10,Tas19}. This closure is based on retaining a finite number (two, in our case) of coefficients in the Hermite-Laguerre expansion of a generalized perturbed distribution function and in simplifying nonlinear terms involving gyroaverage operators, in such a way that the resulting gyrofluid system, if the closure is applied to all species, conserves energy. In Ref. \cite{Tas19} it is shown that such closure also preserves the Hamiltonian character of the parent gyrokinetic model.  It also emerges that that the model (\ref{4fconte}) -(\ref{4fampperp}) is Hamiltonian, although two different closures are adopted for electrons and ions. We also remark that the differences between the two closures emerge not in the evolution equations, but rather in the different treatments of ion and electron terms in the static relations  (\ref{4fqn})-(\ref{4fampperp}).

If one looks at the linear dispersion relation of the parent gyrokinetic model, it turns out that the closures imposed on the moments of higher order with respect to density and parallel momentum fluctuations are compatible with the regime
\beq  \label{condclos}
v_{thi} \ll \frac{\omega}{k_z} \ll v_{the}.
\eeq
 The inequality $v_{thi} \ll \omega/ k_z$ leads to the closure relation for the ions and is compatible with an adiabatic closure for the ion gyrocenter fluid. The inequality $\omega / k_z \ll v_{the}$, on the other hand, concerns the electron fluid and, as above stated, is compatible with an isothermal electron fluid.

The derivation of the two-field model from the four-field model proceeds by first identifying a small parameter which we take to be the square root of the mass ratio
\beq
\delta=(m_e / m_i)^{1/2}.
\eeq
This will allow in particular to neglect most of the electron FLR effects. Corrections of order $\delta$ with respect to the leading order will however be retained in the model. This will make it possible, in particular, to retain electron inertia effects in the parallel electron momentum equation, thus allowing for magnetic reconnection. 

A further assumption is to take
\beq
\frac{\beta_e}{2} =O(\delta) 
\eeq
as $\delta \rightarrow 0$. This will help to make electron FLR terms subdominant with respect to electron inertia terms at finite $\tau$. Also, parallel magnetic perturbations (which, at the leading order, are proportional to $\beta_e$) will be retained as corrections one order smaller in $\delta$.  

We also require that the two-field model retains ion FLR effects, and in particular that it reduces to known reduced models (see Sec. \ref{sec:model}) in the small and large $\tau$ limits.

All these requirements can be met by retaining, from the parent model, (\ref{4fconte}) -(\ref{4fampperp}) only the leading order terms, as well as their first order corrections in $\delta$, on the basis of two scalings. The final model is obtained by considering the union of all the terms present in the reduced models obtained from the two scalings.

More precisely, we first introduce the following \\

{\it \underline {Scaling I}}\\

\begin{align}
&\partial_t=O(\varepsilon \delta^{1/2}), \qquad \pa_x=O(1), \qquad \pa_y=O(1), \qquad \pa_z=O(\varepsilon \delta^{1/2}),    \label{sc11} \\
&\rho_s=O(1), \qquad  d_e^2 = O(\delta), \qquad d_i^2= O(1/\delta), \qquad \frac{\beta_e}{2} =O (\delta), \qquad \tau=O(1)\\
&N_{e,i} = \varepsilon (\delta^{1/2} N_{e,i 0}+\delta^{3/2} N_{e,i 1} + O(\delta^{5/2})),\\
&U_e= \varepsilon (\delta^{1/2} U_{e0}+\delta^{3/2} U_{e1} + O(\delta^{5/2})),  \label{sc14}\\
&U_i=\varepsilon( \delta^2 U_{i0} + \delta^3 U_{i1} +O(\delta^4)),  \label{sc15}\\
&\varphi=\varepsilon(\delta^{1/2}\varphi_0 + \delta^{3/2} \varphi_1 +O(\delta^{5/2})),\\
&\apar=\varepsilon(\delta^{1/2} A_{\parallel 0} + \delta^{3/2} A_{\parallel 1} +O(\delta^{5/2})),  \label{sc17}\\
&B_z=\varepsilon(\delta^{3/2} B_{z 0} + \delta^{5/2} B_{z 1} +O(\delta^{7/2}))  \label{sc18}
\end{align}
and \\

{\it \underline {Scaling II}}
\begin{align}
&\partial_t=O(\varepsilon \delta^{1/2}), \qquad \pa_x=O(1), \qquad \pa_y=O(1), \qquad \pa_z=O(\varepsilon \delta),   \label{sc21} \\
&\rho_s=O(1), \qquad  d_e^2 = O(\delta), \qquad d_i^2= O(1/\delta), \qquad \frac{\beta_e}{2} =O (\delta), \qquad \tau=O(1/\delta),\\
&N_{e,i} = \varepsilon (\delta^{3/2} N_{e,i 0}+\delta^{5/2} N_{e,i 1} + O(\delta^{7/2})),\\
&U_e= \varepsilon (\delta U_{e0}+\delta^{2} U_{e1} + O(\delta^{3})),\\
&U_i=\varepsilon( \delta^2 U_{i0} + \delta^3 U_{i1} +O(\delta^4)),\\
&\varphi=\varepsilon(\delta^{1/2}\varphi_0 + \delta^{3/2} \varphi_1 +O(\delta^{5/2})),\\
&\apar=\varepsilon(\delta A_{\parallel 0} + \delta^{2} A_{\parallel 1} +O(\delta^{3})),\\
&\bpar=\varepsilon(\delta^{3/2} B_{\parallel 0} + \delta^{5/2} B_{\parallel 1} +O(\delta^{7/2})).
\label{sc28}
\end{align}
In Eqs. (\ref{sc11})-(\ref{sc18}) and (\ref{sc21})-(\ref{sc28}) we used the parameter 
\beq
\varepsilon = \frac{L}{L_\parallel} \ll 1,
\eeq
already introduced in Sec. \ref{sec:model}. This parameter accounts for the gyrokinetic ordering but in practice it will play no role in the derivation, because all terms in the evolution equations are of quadratic order in $\varepsilon$ and no further expansion is carried out in this parameter. 

Scalings I and II are consistent (taking into account the different normalization) with those of Ref. \cite{PST18}, although we take this opportunity to point out a misprint in Eq. (18) of Ref.  \cite{PST18}, where the correct expression should be $\apar \sim \partial_z = O(\delta \varepsilon)$.

We begin by deriving a reduced model according to the scaling I, which concerns the case of finite $\tau$. If we insert the relations (\ref{sc11})-(\ref{sc18}) into Eqs. (\ref{4fqn})-(\ref{4fampperp}) we can write
\begin{align}
&\gamui N_i - N_e= \frac{1 - \Gamma_0}{\tau} \frac{\varphi}{\rho_s^2} - (\Gamma_0 - \Gamma_1-1) B_z +O(\varepsilon \delta^{5/2}),  \label{qnsc1}\\
&\lapp \apar =  U_e +O(\varepsilon \delta^2),   \label{ampparsc1}\\
&B_z = -\frac{\beta_e}{2}\left( \tau 2 \gamdi N_i + (\Gamma_0 - \Gamma_1 -1)\frac{\varphi}{\rho_s^2} + (1+2 \tau (\Gamma_0 - \Gamma_1) B_z \right. \nno \\
&\left. +  N_e\right) +O(\varepsilon \delta^{7/2}).  \label{ampperpsc1}
\end{align}
Note that electron FLR corrections are all included in the terms of the form $"O( \, \, )"$, which we refer to as "unspecified quantities". In particular, for a function $f$ assumed of order unity, one has $\gamue f = 2\gamde f=  f + O(\delta^2)$. Therefore, electron FLR corrections are at least of order $\delta^2$ smaller than the dominant contribution in each equation.

If we neglect the unspecified terms in each equation we obtain
\begin{align}
&\gamui N_i - N_e= \frac{1 - \Gamma_0}{\tau} \frac{\varphi}{\rho_s^2} - (\Gamma_0 - \Gamma_1-1) B_z,  \label{qnsc1red}\\
&\lapp \apar =  U_e,   \label{ampparsc1red}\\
&B_z = -\frac{\beta_e}{2}\left( \tau 2 \gamdi N_i + (\Gamma_0 - \Gamma_1 -1)\frac{\varphi}{\rho_s^2} + (1+2 \tau (\Gamma_0 - \Gamma_1)) B_z +  N_e\right).  \label{ampperpsc1red}
\end{align}
 Note, however, that by neglecting the unspecified terms, {\it not all} the corrections which are $\delta^2$ smaller than the dominant contribution have been eliminated. In fact, the whole expansions, in powers of $\delta$ of the fields, are retained. For instance, in Eq. (\ref{ampparsc1red}), using Eqs. (\ref{sc14}) and (\ref{sc17}), one finds
\beq
 \varepsilon\delta^{1/2}\lapp A_{\parallel 0}+\varepsilon\delta^{3/2} \lapp A_{\parallel 1} + O(\varepsilon\delta^{5/2}) = \varepsilon\delta^{1/2}U_{e0} + \varepsilon\delta^{3/2} U_{e1} + O(\varepsilon\delta^{5/2}).
 \eeq
 All terms at the leading order $O(\varepsilon \delta^{1/2})$ and at the next order $O(\varepsilon \delta^{3/2})$ are correctly present. However, terms of order $O(\varepsilon\delta^{5/2})$ are  in principle retained in Eq. (\ref{ampparsc1red}), although the unspecified terms of order $O(\varepsilon \delta^2)$ (which are of lower order) have been neglected. With this approach, therefore, some subdominant terms are retained. More generally, each equation of the final model will turn out to be correct at the leading order, at the next order in $\delta$ but will also contain some subdominant higher order terms. It is therefore assumed that such subdominant terms do not play a major role in the dynamics. As discussed in Ref. \cite{GTLPS23}, this assumption fails, as expected, at sufficiently small scales, typically of the order of the electron Larmor radius. On the other hand, the advantage of this approach is that it is significantly simpler than the standard consistent approach, where equations are solved order by order and only terms of the same order are retained in the final model equations.
 
 Inserting the expansions (\ref{sc11})-(\ref{sc18}) into the evolution equations (\ref{4fconte})-(\ref{4fmomi}) yields
\begin{align}
&\frac{\pa N_e}{\pa t}+ [ \varphi - \rho_s^2  B_z , N_e] - [ \apar , U_e]+\frac{\pa U_e}{\pa z}+O(\varepsilon^2 \delta^{3})=0,  \label{4fcontesc1}\\
&\frac{\pa}{\pa t}( \apar - d_e^2 U_e)+ [ \varphi - \rho_s^2  B_z ,  \apar - d_e^2 U_e] +\rho_s^2 [  \apar , N_e]  \nno \\
& +\frac{\pa }{\pa z}( \varphi - \rho_s^2 ( B_z + N_e))+O(\varepsilon^2 \delta^{3})=0,  \label{4fmomesc1}\\
&\frac{\pa N_i}{\pa t}+ [\gamui \varphi + \tau \rho_s^2 2 \gamdi B_z , N_i] +O(\varepsilon^2 \delta^{5/2})=0,  \label{4fcontisc1}\\
&\frac{\pa}{\pa t}(\gamui \apar + d_i^2 U_i) + [\gamui \varphi + \tau \rho_s^2 2 \gamdi B_z , \gamui \apar] + [\gamui \varphi, d_i^2 U_i] -\tau \rho_s^2 [ \gamui \apar , N_i]  \nno \\
& +\frac{\pa }{\pa z}(\gamui \varphi + \rho_s^2 (\tau 2 \gamdi B_z + N_i))+O(\varepsilon^2 \delta^{5/2})=0.  \label{4fmomisc1}
\end{align}
Note that, in Eq. (\ref{4fmomesc1}), the term $[\rho_s^2 B_z, d_e^2 U_e]$, which is of order $\varepsilon^2 \delta^3$, has been explicitly retained, although it is $\delta^2$ smaller than the leading order terms in the equation. It can be verified a posteriori, that, in the presence of such term, the model possesses a Hamiltonian structure, which is a property that we require for our model.

Again, neglecting the unspecified terms, Eqs. (\ref{4fcontesc1})-(\ref{4fmomisc1}) can be written as
\begin{align}
&\frac{\pa N_e}{\pa t}+ [ \varphi - \rho_s^2  B_z , N_e] + \nabla_\| U_e=0,  \label{4fcontesc1red}\\
&\frac{\pa}{\pa t}( \apar - d_e^2 U_e)- [ \varphi - \rho_s^2  B_z ,  d_e^2 U_e] +\nabla_\| (\varphi -\rho_s^2  N_e - \rho_s^2 B_z)  =0,  \label{4fmomesc1red}\\
&\frac{\pa N_i}{\pa t}+ [\gamui \varphi + \tau \rho_s^2 2 \gamdi B_z , N_i] =0,  \label{4fcontisc1red}\\
&\frac{\pa}{\pa t}(\gamui \apar + d_i^2 U_i) + [\gamui \varphi + \tau \rho_s^2 2 \gamdi B_z , \gamui \apar] + [\gamui \varphi, d_i^2 U_i] -\tau \rho_s^2 [ \gamui \apar , N_i]  \nno \\
& +\frac{\pa }{\pa z}(\gamui \varphi + \rho_s^2 (\tau 2 \gamdi B_z + N_i))=0.  \label{4fmomisc1red}
\end{align}
We can then take 
\beq
N_i=0
\eeq
as solution of Eq. (\ref{4fcontisc1red}). This shows a posteriori, that in scaling I, one could have alternatively assumed $N_i=\varepsilon (\delta^{5/2} N_{i0} + \delta^{7/2} N_{i1} + O(\delta^{9/2}))$. Either way, $N_i$ becomes negligible in Eqs. (\ref{qnsc1red}) and (\ref{ampperpsc1red}). Also, $U_i$ has been neglected in the approximated parallel Amp\`ere's law   (\ref{ampparsc1red}). Thanks to these decouplings, from Eqs. (\ref{4fcontesc1red}), (\ref{4fmomesc1red}), (\ref{ampparsc1red}), (\ref{ampperpsc1red}), one obtains a closed two-field system given by 
 \begin{align}
&\frac{\pa N_e}{\pa t}+ [ \varphi - \rho_s^2  B_z , N_e] + \nabla_\| \lapp \apar=0,  \label{2fcontesc1fin}\\
&\frac{\pa}{\pa t}( \apar - d_e^2 \lapp \apar)- [ \varphi - \rho_s^2  B_z ,  d_e^2 \lapp \apar] +\nabla_\| (\varphi -\rho_s^2  N_e - \rho_s^2 B_z)  =0,  \label{2fmomesc1fin}\\
& N_e= \frac{ \Gamma_0 -1}{\tau} \frac{\varphi}{\rho_s^2} - (1- \Gamma_0 + \Gamma_1) B_z,  \label{qnsc1fin}\\
&B_z = -\frac{\beta_e}{2}\left( (\Gamma_0 - \Gamma_1 -1)\frac{\varphi}{\rho_s^2} + (1+2 \tau (\Gamma_0 - \Gamma_1)) B_z +  N_e\right), \label{ampperpsc1fin}
\end{align}
where we also made use of Eq. (\ref{ampparsc1red}) to replace $U_e$ in favour of $\lapp \apar$.  

The same procedure is then applied to the parent four-field model with scaling II. We just point out that, in the case of scaling II, which deals with $\tau \gg1$, the decoupling of $N_i$ and $U_i$ occurs due to the ion gyroaverage operators, which damp $N_i$ and $U_i$ exponentially.

The two-field model resulting from scaling II reads
 \begin{align}
&\frac{\pa N_e}{\pa t}+ [ \varphi - \rho_s^2  B_z , N_e] + \nabla_\| \lapp \apar=0,  \label{2fcontesc2fin}\\
&\frac{\pa}{\pa t}( \apar - d_e^2 \lapp \apar)- [ \varphi - \rho_s^2  B_z ,  d_e^2 \lapp \apar] +\nabla_\| (\varphi -\rho_s^2  N_e - \rho_s^2 B_z)  =0,  \label{2fmomesc2fin}\\
& N_e= \left(-\frac{1}{\tau}+\bdep \lapp\right) \frac{\varphi}{\rho_s^2} - B_z,  \label{qnsc2fin}\\
&B_z = -\frac{\beta_e}{2}\left(  -\frac{\varphi}{\rho_s^2} +  B_z +  N_e\right).\label{ampperpsc2fin}
\end{align}
The evolution equations look identical to those obtained from scaling I, although formally, the leading order terms in Eqs. (\ref{2fcontesc2fin}) and (\ref{2fmomesc2fin}) are $O(\varepsilon^2 \delta^2)$ and $O(\varepsilon^2 \delta^3/2)$, respectively, whereas leading order terms in Eqs. (\ref{2fcontesc1fin}) and (\ref{2fmomesc1fin}) are both $O(\varepsilon^2 \delta)$. We emphasize the presence of the only electron FLR correction appearing in our model, and given by the term $(\beta/2)d_e^2 \lapp \varphi/\ \rho_s^2$ in Eq. (\ref{qnsc2fin}). This term emerges for $\tau \gg 1$ as correction of order $\delta$ smaller, and is thus retained in the quasi-neutrality relation. The importance of this term for vortex formation has been discussed in Sec. \ref{sec:reconnection}.

The final two-field gyrofluid model consists of the two evolution equations 
\begin{align}
&\frac{\pa N_e}{\pa t}+ [ \varphi - \rho_s^2  B_z , N_e] + \nabla_\| \lapp \apar=0,  \label{2fcontefina}\\
&\frac{\pa}{\pa t}( 1 - d_e^2 \lapp)\apar- [ \varphi - \rho_s^2  B_z ,  d_e^2 \lapp \apar] +\nabla_\| (\varphi -\rho_s^2  N_e - \rho_s^2 B_z)  =0,  \label{2fmomefina}
\end{align}
obtained from both scalings, and given by Eqs. (\ref{2fcontesc1fin})-(\ref{2fmomesc1fin}) or, equivalently, by Eqs. (\ref{2fcontesc2fin})-(\ref{2fmomesc2fin}). These equations correspond indeed to Eqs. (\ref{contfin})-(\ref{ohmfin}).

These two evolution equations are complemented by two static relations obtained taking the union of all the terms in the static relations obtained from the two scalings, i.e. Eqs. (\ref{qnsc1fin})-(\ref{ampperpsc1fin}) and (\ref{qnsc2fin})-(\ref{ampperpsc2fin}). In this way, one can recover the appropriate limit of the static relations, by applying the corresponding scaling. This procedure yields
\begin{align}
& N_e =   \left(\frac{\Gamma_{0} -1}{\tau}+\bdep \lapp\right)\frac{\varphi}{\rho_s^2}-(1 - \Gamma_{0}  +\Gamma_{1}) B_z,  \label{qnprovv}\\
& B_z = -\frac{\beta_e}{2}\left( (\Gamma_0 - \Gamma_1 -1)\frac{\varphi}{\rho_s^2} + (1+2 \tau (\Gamma_0 - \Gamma_1)) B_z +  N_e\right).  \label{ampperpprovv}
\end{align}
Making use of Eq. (\ref{qnprovv}), from Eq. (\ref{ampperpprovv}) one can obtain
\beq
\left( \frac{2}{\beta_e} + (1 + 2 \tau) (\Gamma_0  -\Gamma_1 )\right) \bpar=\left(1 - \frac{\Gamma_0  -1}{\tau} -\Gamma_0  +\Gamma_1  \right) \frac{\varphi}{\rho_s^2}.  \label{ampperpprovv2}
\eeq
Note that, in the right-hand side of Eq. (\ref{ampperpprovv2}), a term $-(\beta_e/2) d_e^2 \lapp \varphi / \rho_s$ has been neglected. Within our approach this is legitimate, because such term would only contribute to terms which are of order $\delta^2$ smaller than the leading terms in each equation. As above explained, our model is not accurate at the level of such terms. 

We can now summarize by pointing out that Eqs. (\ref{2fcontefina}), (\ref{2fmomefina}), (\ref{qnprovv}) and (\ref{ampperpprovv2}) correspond to Eqs. (\ref{contfin}), (\ref{ohmfin}), (\ref{qnfin}) and (\ref{ampperpfin}), respectively. This concludes the derivation of the model.

\section*{Acknowledgments}
This work was performed using HPC resources from GENCI - CINES/TGCC (Grants 2020 - A0090407042 and 2021 - A0110407042), and using HPC resources from GENCI-TGCC (Grant 2022 - A0130413794).
Part of the computations have also been done on the “Mesocentre SIGAMM” machine, hosted by Observatoire de la Côte d’Azur.




\bibliographystyle{elsarticle-num-names}
\bibliography{biblio}

\end{document}